\documentclass{emulateapj}

\usepackage{natbib}
\usepackage{amsmath}
\usepackage{ mathrsfs }
\usepackage{subfigure}
\usepackage{epstopdf}

\long\def\symbolfootnote[#1]#2{\begingroup%
\def\thefootnote{\fnsymbol{footnote}}\footnote[#1]{#2}\endgroup}

\newcommand{\beq}{\begin{equation}}
\newcommand{\eeq}{\end{equation}}
\newcommand{\bea}{\begin{eqnarray}}
\newcommand{\eea}{\end{eqnarray}}

\newcommand{\vphi}{{v_\varphi}}
\newcommand{\Lphi}{{\ell}_\varphi}
\newcommand{\Boxsize}{{\mathscr{B}} }
\newcommand{\Resolution}{\Re}
\newcommand{\Diffusion}{{\cal D}}
\newcommand{\vesc}{v_{\rm esc}}

\begin{document}
\title{Oblique Shock Breakout in Supernovae and Gamma-Ray Bursts: II. Numerical Solutions For Non-Relativistic Pattern Speeds}
\slugcomment{Submitted to ApJ}

\shorttitle{Oblique Shock Breakout in SNe}
\shortauthors{Salbi, P. et al}

\author{Pegah Salbi,\altaffilmark{1} Christopher D. Matzner,\altaffilmark{1} Stephen Ro,\altaffilmark{1} and Yuri Levin\altaffilmark{2}}

\altaffiltext{1}{Department of Astronomy and Astrophysics, University of Toronto, 50 St. George Street, Toronto, Ontario, M5S 3H4, Canada}
\altaffiltext{2}{Monash Center for Astrophysics, Monash University, Clayton, VIC 3800, Australia}
\email{salbi@astro.utoronto.ca}

\begin{abstract}
Non-spherical explosions develop non-radial flows as the pattern of shock emergence progresses across the stellar surface.  In supernovae these flows can limit ejecta speeds, stifle shock breakout emission, and cause collisions outside the star.  Similar phenomena occur in stellar and planetary collisions, tidal disruption events, accretion-induced collapses, and propagating detonations.  
We present two-dimensional, nested-grid Athena simulations of non-radial shock emergence in a frame comoving with the breakout pattern, focusing on the adiabatic, non-relativistic limit in a plane stratified envelope. 
We set boundary conditions using a known self-similar solution and explore the role of box size and resolution on the result.  The shock front curves toward the stellar surface, and exhibits a kink from which weak discontinuities originate.  Flow around the point of shock emergence is neither perfectly steady nor self-similar.  Waves and vortices, which are not predominantly due to grid effects, emanate from this region.  The post-shock flow is deflected along the stellar surface, and its pressure disturbs the stellar atmosphere upstream of the emerging shock.  We use the numerical results and their analytical limits to predict the effects of radiation transfer and gravity, which are not included in our simulations.  
\end{abstract}


\section{Introduction}\label{S:Intro}
The arrival of a normal shock at the surface of an exploding star is associated with a flash of radiation and the release of the fastest stellar ejecta, which can then interact with circumstellar material in the earliest phase of a supernova remnant.  All of these phenomena result from shock acceleration in the steeply declining density profile of the outer stellar envelope: a whip-like motion described by the similarity solution of \citet{1956SPhD....1..223G}, \citet{sakurai60}, and \citet{1999ApJ...510..379M}.  The outcome is very different, however, when the explosion shock is not aligned with the density gradient, so that non-radial flows develop (\citealt{2013ApJ...779...60M}, hereafter Paper 1).   In this case, the pattern of shock emergence moves across the star's surface at some speed $\vphi$.   
A natural length scale is the `obliquity' depth $\Lphi$ at which the shock front is expected to move outward at $\vphi$. So long as radiation is trapped at this depth, motions become strongly non-radial. The shock and ejecta velocity are both limited (to $\vphi$ and $2\vphi$, respectively, in the star's rest frame) and the ejecta spray is expected to suppress the escape of photons which would otherwise contribute to the photon flash. In the aftermath, it is possible that non-radial ejecta will collide outside the star. 

Fully characterizing this behavior requires numerical simulations, as the analytical arguments presented in Paper 1 leave many questions unresolved.  The detailed shape of the shock front, the shape and terminal angle of the outflowing stream lines, and the motions of matter and energy in the emerging flow are all to be determined.   Moreover, Paper 1 showed that the flow immediately around the point of breakout cannot be both steady and self-similar in the comoving frame: it could be oscillatory, or the outflow could interact with the star to spoil the apparent self-similarity.  We construct numerical experiments to address these questions and glean additional insights into the nature of oblique breakouts. 

\section{Physical problem and numerical implementation} \label{S:Problem} 

We simulate only a single, asymptotic version of the oblique breakout problem, which was also a focus of attention in Paper 1.   We ignore the diffusion of radiation, which is valid if $\Lphi$ is below the depth at which diffusion becomes important.  We also ignore the curvature of the stellar atmosphere and any curvature of the shock front: a necessary condition for this to be valid is that $\Lphi$ is much less than the stellar radius.   We can thus consider the problem in a frame of reference co-moving with the breakout pattern, and we can focus our attention on the two-dimensional, stationary flow of an adiabatic gas, separated by a shock front from  plane-stratified upstream fluid.  Furthermore, we specialize to the case where the post-shock flow is radiation dominated (adiabatic index $\gamma=4/3$) and consider a cold polytropic atmosphere (index $n=3$ or $\gamma_p = 4/3$) for the pre-shock matter.    Finally, we assume that the motions are purely non-relativistic, and that stellar gravity and hydrostatic pressure are both negligible.   This combined limit is most applicable to aspherical explosions of compact stars, as was demonstrated in section 5 of Paper 1, but it remains a useful analogue in all cases where obliquity affects the flow.  We shall return to consider the physics we have limited in later sections (radiation transfer in \S \ref{S:OpticalDepth}; gravity in \S \ref{S:Gravity}; relativity and finite depth in Table \ref{Table:EjectaDistribution}). 

These choices have several advantages.  First, as discussed in Paper 1, the shock front is effectively horizontal far below a depth of $\Lphi$.  Because we ignore stellar curvature, all the streamlines which originate from this zone must approach the known planar, self-silmilar solution.  We use this fact when determining both our outer boundary conditions and our initial conditions, as we describe more fully below.  The pattern speed, obliquity scale, and density at the obliquity scale set our code units: $\Lphi = \rho_\varphi=\vphi=1$. 

Second, by ignoring the length scales associated with curvature of the star or the shock front, as well as the length scale on which radiation diffusion becomes important, we are left with only one physical length scale, $\Lphi$, against which to judge our numerical parameters.   If we conduct our simulation in a square box of width $L_x$ whose square grid has a finest scale $\Delta x_{\rm min}$, its outcome can be characterized by the box-size parameter 
\[ \Boxsize = L_x/\Lphi\] and the resolution  parameter \[\Resolution = \Lphi/\Delta x_{\rm min}. \]

Third, we expect that once the outcome is normalized to its natural units of length, time, and density ($\Lphi$, $\Lphi/\vphi$, and $\rho_\varphi$, respectively), it is uniquely determined (at least in a time-averaged sense) by the post-shock adiabatic index $\gamma$ and the polytropic index $\gamma_p$.  (We address only the case $\gamma=\gamma_p=4/3$ in our simulations.)  So long as our boundary conditions are consistent with this flow in the limit of an infinitely large box, this expectation gives us confidence that the results (apart from an initial transient) must limit to the definite physical solution as $(\Boxsize,\Resolution)\rightarrow \infty$.  

\subsection{Code and code test} \label{S:Code_Test} 
\begin{figure}
\includegraphics[scale=0.45]{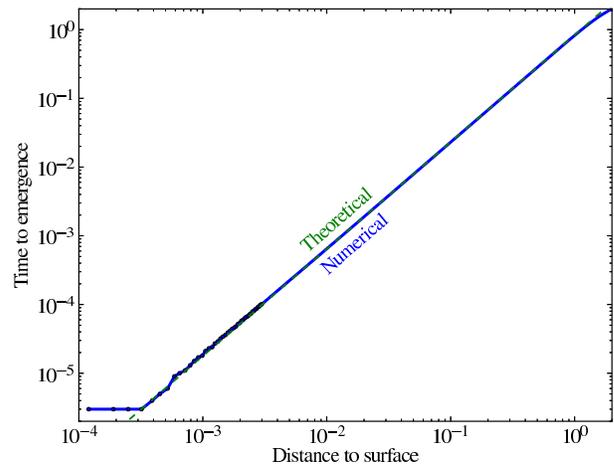}
\caption{Shock velocity as a function of fluid density in a one-dimensional test problem (\S \ref{S:Code_Test}).  A stellar atmosphere of depth two arbitrary units is resolved by $3\times10^4$ grid cells, of which the outer 75 are plotted as dots, and a shock is launched from the innermost region (right-hand side in this figure).  Plotted is the time to shock emergence, calculated using the time of greatest compression in each cell, against the distance to the surface.  Finite resolution affects shock propagation within about ten cells from the surface,  and our initialization affects it for the deepest matter, but for depths between $0.003$ and 0.7 in these units, the shock acceleration law agrees with the self-similar theory of \citeauthor{sakurai60} (\citeyear{sakurai60}; dashed line) within 0.4\%.} 
\label{fig:v_sh_rho}
\end{figure}

We employ the Athena magneto-hydrodynamics code \citep{2008ApJS..178..137S} with magnetic fields turned off.  
Although this code is thoroughly tested on other problems, we wished to check its performance in the context of an accelerating shock.  We therefore set up a one-dimensional problem in which a shock front moves  through cold gas away from a high-pressure region (a region of high initial temperature, next to a reflecting boundary condition), descending an initial density distribution $\rho_0(y) \propto \max[(-y)^3, 0]$ (plus a floor density $10^{-14}$ times lower than the minimum for $y<0$), where, as in Paper 1, the coordinate $y$ is an altitude.   The adiabatic index is $\gamma=4/3$.  The results of this test are presented in Figure~\ref{fig:v_sh_rho}, where we plot the 
time to shock emergence against the initial depth.   
The results adhere within 0.4\% to the theoretical power law $(t_{\rm breakout}-t)\propto |y|^{1.55724}$ derived by \citet{2013ApJ...773...79R} using the self-similar theory of \citet{sakurai60}.  

For our two-dimensional simulations Athena was compiled with Message Passing Interface (MPI) and Static Mesh Refinement (SMR) enabled. We used a Harten-Lax-van Leer-Contact (HLLC) Riemann solver with second-order reconstruction, with a Corner Transport Upwind (CTU) unsplit integrator, and Courant number of 0.8.  The computations were performed on 256 cores of the GPC cluster at the SciNet facility located at the University of Toronto \citep{scinet}.

We employ a sequence of three nested grids, in which one grid (of resolution up to $4096^2$) is nested within an identical grid scaled up by a factor of two in size, and this in turn is nested within a third identical grid, another factor of two larger, for an effective linear dynamical range $\Boxsize\Resolution=16,384$.    (A code error prevented us from using even more than three resolution levels.)  

\subsection{Initial and boundary conditions} \label{S:Inits_and_BCs}
We appeal to the known planar, self-similar solution, and the fact that the flow limits to this solution for large initial depths ($y_0\ll -\Lphi$) to set our initial and boundary conditions.   In this planar solution, profiles of velocity, density, and pressure behave self-similarly, adhering to fixed functional forms which scale in amplitude and length scale as the shock approaches the stellar surface.  To use it, we translate this time-dependent, one-dimensional flow into two dimensions.  This involves several steps.  First, we match the upstream density in the one-dimensional solution to the depth-dependent initial density $\rho_0(y_0)$ in the 2D grid.  Then we match the shock strength, ensuring $v_s = \vphi$ where $y_0=-\Lphi$; this enforces the definition of $\Lphi$ from Paper 1.  Third, we associate each time in the 1D solution with a location in the 2D flow: $(y,t') \rightarrow (y,x)$ where $x=\vphi t$, and copy all of the fluid variables from the 1D solution to the 2D simulation volume.   (This step is performed using a seventh-order polynomial fit to the self-similar solution, accurate to a few tenths of a percent in each variable, for computational ease.)  Matter ahead of the shock front is assigned its initial density, $\rho(x,y)=\rho_0(y) =  (|y|/\Lphi)^3\rho_\varphi$ for $y<0$). For $y>0$ we set the density equal to a low value (a lower density than for any $y<0$), and pressure to a floor value of $10^{-10}$ in code units. In other words, we ignore the portion of the 1D solution in which ejecta flies away ($y>0$, $t'>0$); this material does not contribute to the final solution, and introduces very strong gradients which lead to numerical problems.  The pressure floor is applied also to the pre-shock fluid ($y<0$).  Finally, we set the horizontal velocity: $v_x=\vphi$ everywhere: this enforces the inflow and outflow of matter across the left and right grid boundaries, respectively.   These initial conditions are depicted in the first panel of Figure \ref{fig:FiducialRun}. 

Except for our treatment of the $y>0$ material, these `self-similar' initial conditions correspond to a scenario in which the fluid responds only to the vertical component of its pressure gradient, as it does in the 1D solution.  For matter which originates deep within the star ($y_0\ll -\Lphi$) this approximation is valid, as its shock normal and the post-shock pressure gradients are indeed vertical.  For matter which originates in the oblique zone ($y_0\gtrsim -\Lphi$) it is far from correct, and we expect a transient period of readjustment, lasting a few flow times $\Lphi/\vphi$,  before it can establish a self-consistent stationary state.     Because our runs are limited to finite volume (finite $\Boxsize$), we cannot perfectly reproduce the ideal solution. 

These initial conditions also provide boundary conditions for the remainder of the simulation.  The bottom boundary, and the inflow region ($y<0$) of the left boundary are set to enforce inflow: the initial fluid variables are fixed (in a ghost region) for all time.  All other boundaries are assigned outflow conditions, in which accelerations at the boundary are calculated using a linear extrapolation of the fluid variables.    The final stationary solution has supersonic outflow in all the outflow regions of the grid boundary, so our choice is self-consistent.  

In the project's initial stages we used a small wedge of very high-density zones to launch an oblique shock upward through the density gradient.   Although the flow very close to the stellar surface was similar in these runs to what we saw later with self-similar boundary conditions, we deemed this procedure unsatisfactory.   A sequence of runs with different wedge parameters (height, angle, overdensity factor) revealed that a large portion of the simulation volume was affected by these parameter choices. Mixing of matter between the wedge and the stellar envelope, and the dynamical reaction of the wedge, complicated the analysis.  Furthermore high density contrasts between the wedge and the envelope induced frequent crashes of the code.  For these reasons we report only on runs with the self-similar boundary conditions described above, which have the benefit that they become exact in the limit $\Boxsize\rightarrow \infty$.

\section{Results - Fiducial simulation} \label{S:FiducialCaseResults} 

\begin{figure}[ht]
\centering
\subfigure[Initial conditions and refinement zones]{\includegraphics[scale=0.18]{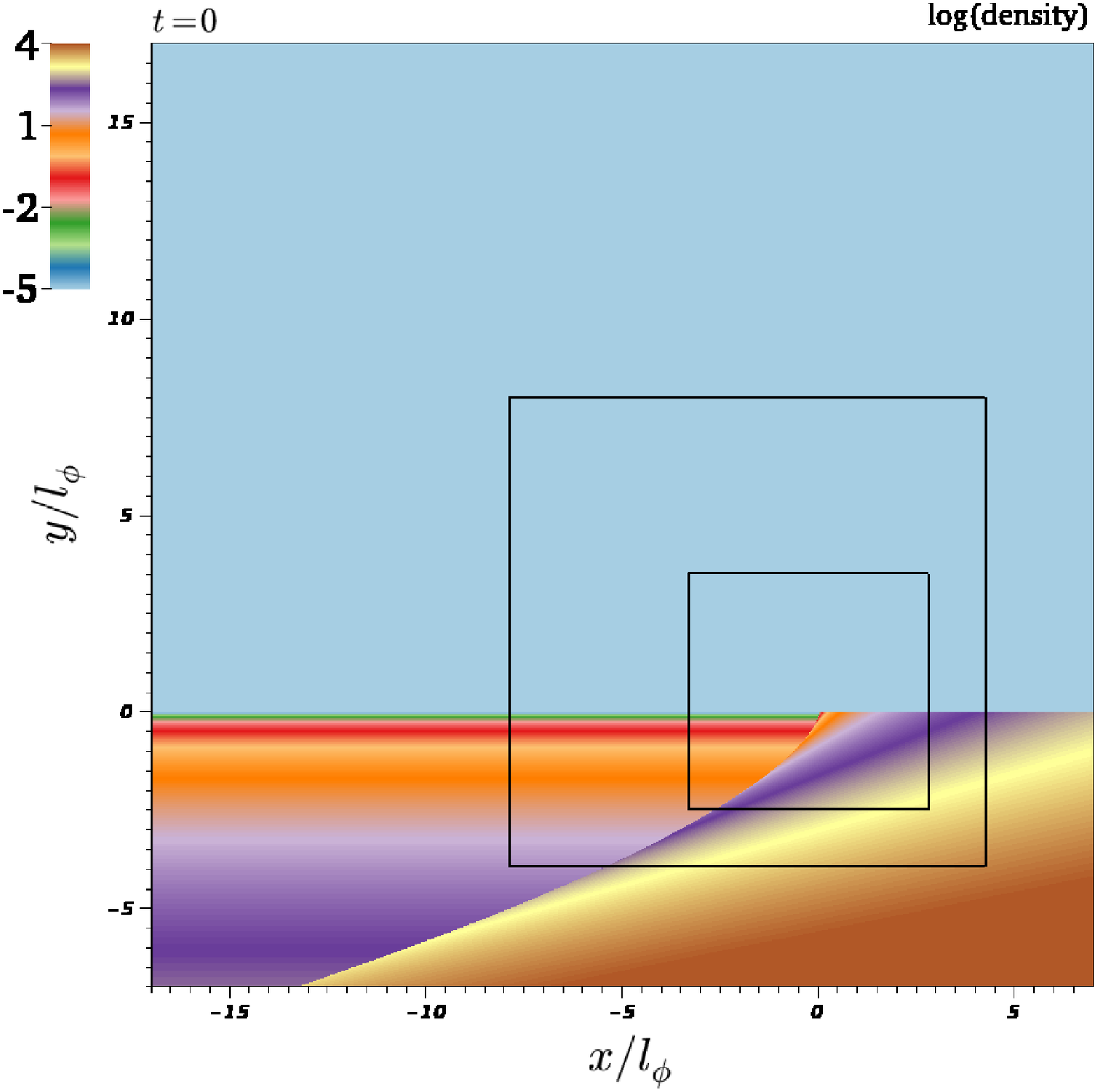}
}
\quad
\subfigure[Transient evolution]{\includegraphics[scale=0.19]{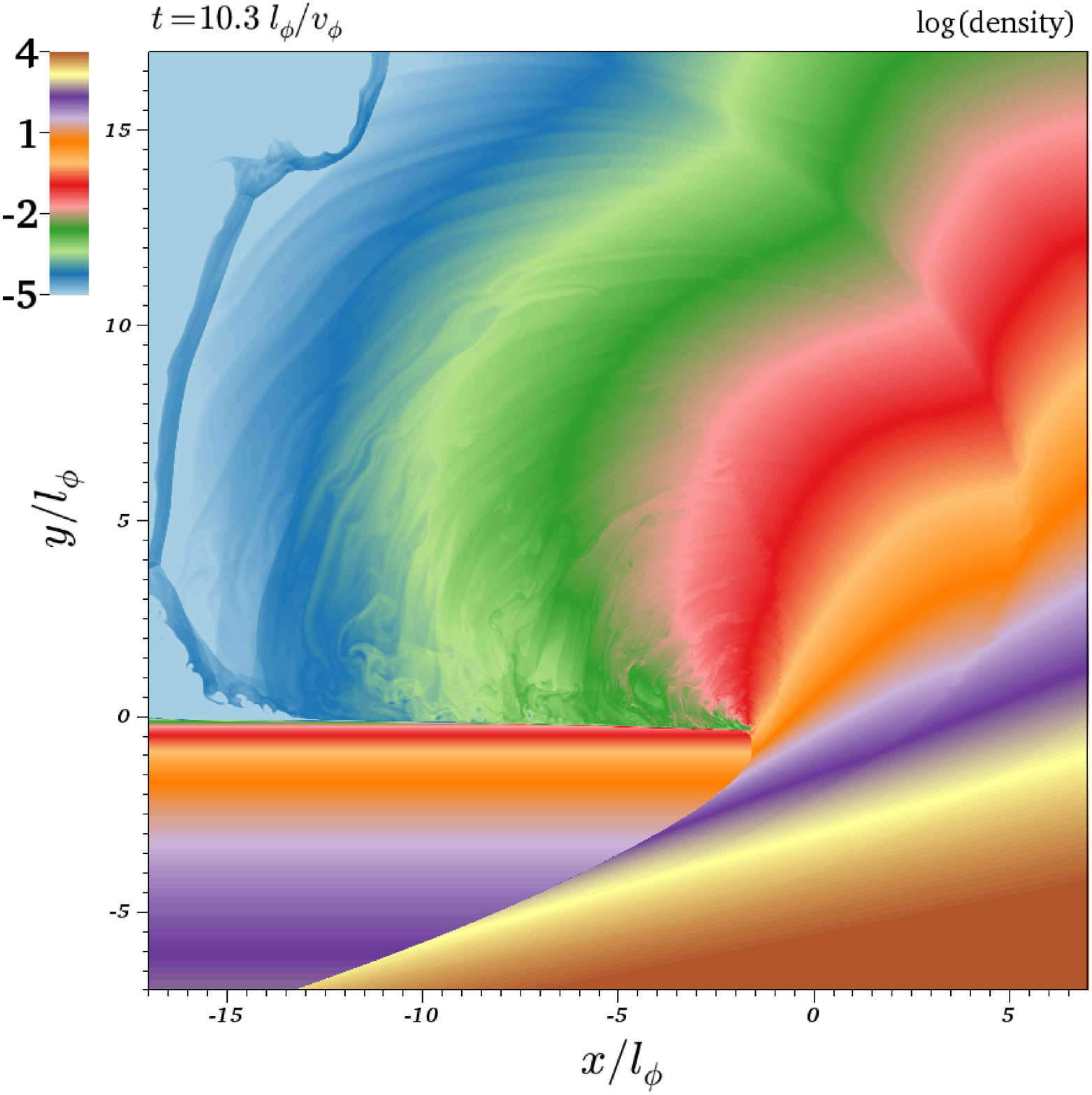}
} 
\quad
\subfigure[Final stationary state]{ \includegraphics[scale=0.19]{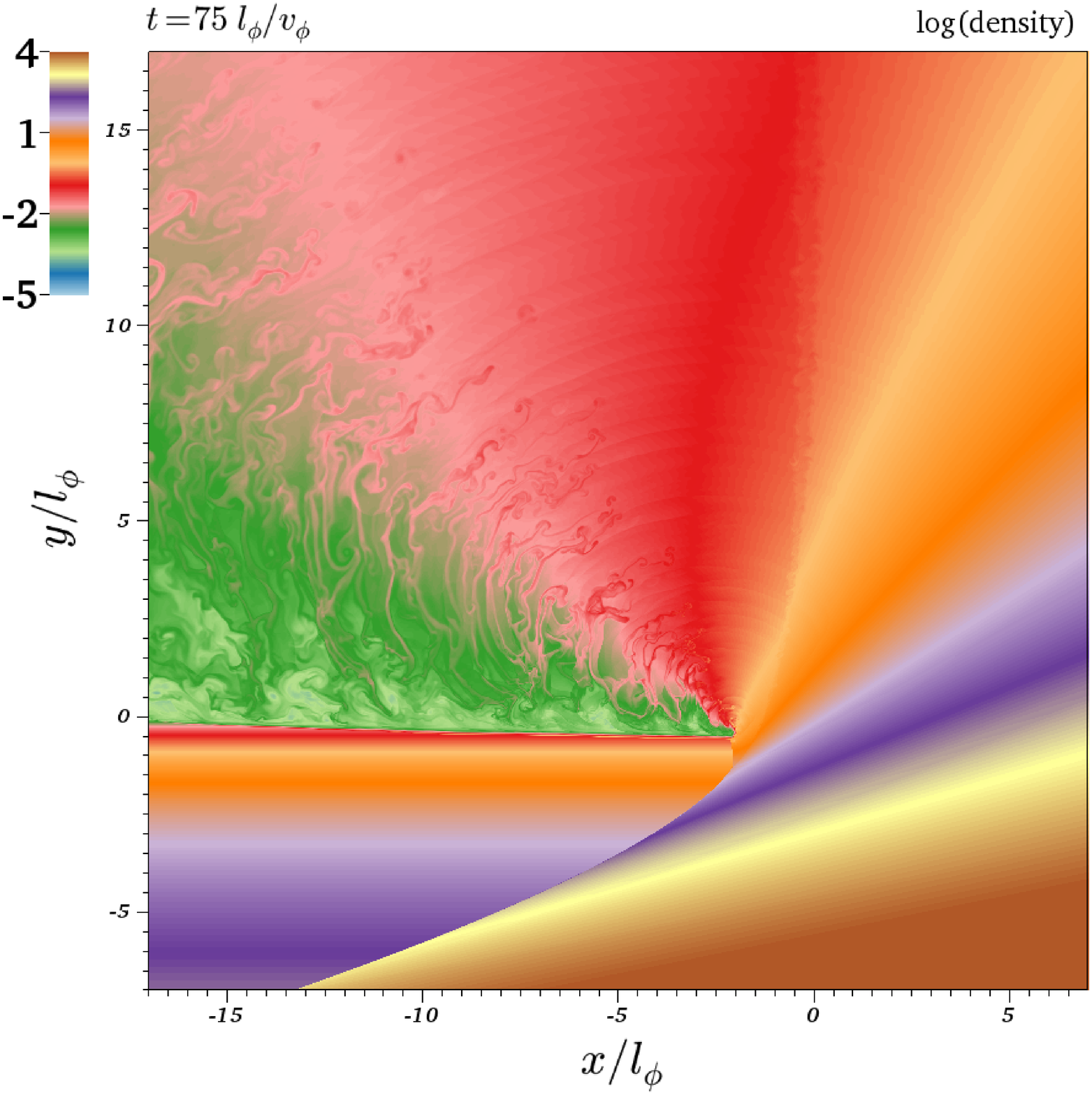}
}
\caption{Snapshots of density distribution in the fiducial run.  The simulation is performed in a frame comoving with the pattern of shock emergence as it traverses the stellar surface. Matter flows from the left and across the shock front before accelerating outward.  Note: in all figures except Figure \ref{Fig:Finest_resolution}, only the coarsest grid level is rendered. The boxes in the top panel denote the location of the finer grids.}  \label{fig:FiducialRun} 
\end{figure}

Figure \ref{fig:FiducialRun}
depicts the progression of the density distribution in our fiducial run, for which $\Resolution = 683$, $\Boxsize=24$, the largest grid extends over $-7\Lphi<(x,y)<17\Lphi$, and the finest grid extends over $-3.5\Lphi<x<2.5\Lphi$, $-2.5\Lphi<y<3.5\Lphi$. Because this run represents the largest box and the highest resolution achieved in our study, we wish to describe its features before examining the effects of resolution (\S \ref{SS:Resolution}) and box size (\S \ref{SS:Boxsize}) on the results. 

As expected, we see a period of transient evolution from the initial state.  Post-shock matter flies upward, establishing an outflow above the stellar surface, and is deflected upstream (to negative $x$) by its internal pressure gradients.   The final panel of Figure  \ref{fig:FiducialRun} represents the statistically stationary final state, in which only short-period oscillations persist.   Figures \ref{Fig:ShockFit}, \ref{Fig:compression}, \ref{Fig:vorticity}, \ref{Fig:pressure}, and \ref{Fig:entropy} depict the shock structure, compression rate, specific vorticity, pressure, and entropy, respectively, of the final stationary state, and figure \ref{Fig:Finest_resolution} provides a magnified view of the entropy structure in the highest-resolution subgrid, immediately around the breakout region.  We do not plot velocity vectors, but we note that, in the shock's frame, they are almost constant in magnitude (except in the immediate post-shock region where the shock is vertical) and they are parallel to contours of entropy (except where vortices develop).  They are therefore very similar to the velocity field predicted in Figure 1 of Paper 1. 

In the following subsections we comment on several aspects of the final stationary state.

\subsection{Shock structure}  \label{SS:shock_structure} 

During the emergence of an aspherical supernova shock, part of the stellar surface has been shocked while another part has not.  As a consequence of its acceleration, the explosion shock (which divides these regions) curves outward, with its normal vector becoming more non-radial, as it approaches the stellar surface.   Several observations made in Paper 1 form our expectations on the shape of the shock front.    

\begin{figure}
\includegraphics[scale=0.5]{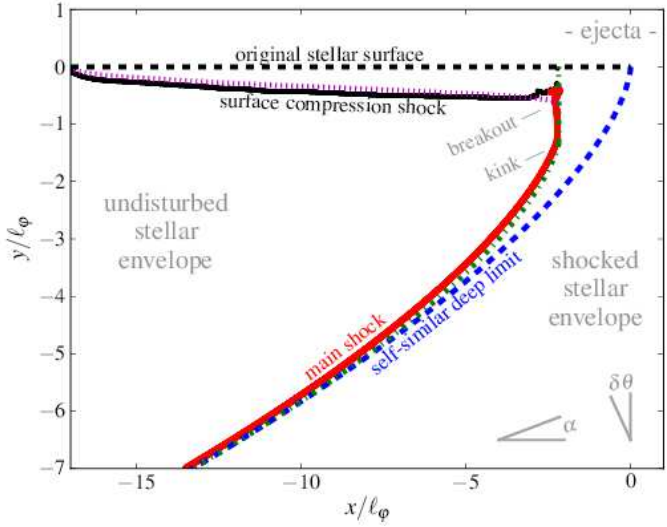}
\caption{Primary shock locus (red line) at $t=75\,t_\varphi$ in the fiducial run, compared with the asymptotic power-law form used to set boundary conditions (eq. \ref{Shock_Shape_in_Deep_Zone_of_Steady_Flow}, blue dash-dot line) and with the approximate model of \citeauthor{1964PThPh..32..207I} (1964, green dash-dot line).  The two free parameters of this model are set by matching the value and slope to eq.~(\ref{Shock_Shape_in_Deep_Zone_of_Steady_Flow}) at $y=-6.48\Lphi$, a choice which optimizes its match to the simulation.  The simulation shock experiences a kink at $(x,y)=(-2.11, -1.32)\Lphi$, whereas the model's kink is at $(-2.25,-1.34)\Lphi$.   As in Paper I, we extend the model vertically to the stellar surface from this kink location.   Comparing the original stellar surface (black dashed line) and the lower boundary of the region disturbed by outflow (black solid line) illuminates the ejecta-envelope interaction discussed in \S \ref{SS:ejecta_surface_interaction}.  The  purple dotted line depicts equation (\ref{eq:shock-depression}), which approximates this interaction.  The comoving-frame and rest-frame deflection angles ($\alpha$ and $\delta \theta$, respectively) are depicted for clarity. }
\label{Fig:ShockFit} 
\end{figure}

First, because we limit ourselves to adiabatic flow and neglect stellar curvature, the shock front will evolve self-similarly while it is normal, i.e., while it is much deeper than the obliquity scale.   This fact implies a limiting form for the locus of points $(x_s,y_s)$ which define the shock surface, which indeed we have already imposed in our boundary conditions.   In particular, our definitions imply 
\beq \label{Shock_Shape_in_Deep_Zone_of_Steady_Flow}
\left(-y_s\over \Lphi\right)^{\lambda+1} \rightarrow (\lambda+1){x_s - x_{s0}\over \Lphi} ~~~{\rm for}~ y_s\ll-\Lphi, 
\eeq 
where $\lambda = 0.5574$ for  $n=3$ and $\gamma=4/3$, 
and our initial and boundary conditions correspond to the choice $x_{s0}=0$.  (This choice differs from Paper 1, where we enforced $x=0$ at the point where the shock meets the surface.  The two definitions are not equivalent, because the flow connects the self-similar offset $x_{s0}$ to the point of shock emergence in a definite way.)  This deep self-similar limit is plotted as a blue dashed line in Figure \ref{Fig:ShockFit}.

Second, the theory of \citet{1964PThPh..32..207I} predicts a definite functional form for the shock front (up to a scaling factor) as  its normal vector curves away from the vertical.    A key feature of this prediction is that the shock angle cannot evolve continuously past a specific value (in the theory, this is when the shock tangent is 20.7$^\circ$ from the vertical).  If the shock is to reach the surface, it must experience a kink, then travel to the surface at a set angle.   However, this theory is only approximate. Its predictions are therefore tentative, as is clear from the fact that it is not consistent with the self-similar evolution in the deep limit ($y_s \ll -\Lphi$).   Paper 1 hypothesized that the shock transitions smoothly from its deep limit to the form predicted by \citeauthor{1964PThPh..32..207I} at some reference depth. 

Third, the similarity analysis presented in Paper 1 makes a very weak prediction that the shock should reach the surface vertically (with it a horizontal normal vector).   This analysis provides an analytical form for the flow structure about the point of breakout, but it is not physical (for astrophysical values of $\gamma$ and $\gamma_p$) because it requires the sine of the angle between the shock normal and the vertical to exceed unity by a small amount ($\sin \alpha_s >1$).   While this primarily implies that the surface flow is either not steady or not self-similar (or both), it nevertheless suggests that, if the self-similar solution guides the flow in any way, its shock normal will be as close as possible to the self-similar condition.  The closest physical solution is a vertical shock front ($\sin \alpha_s = 1$).  

\begin{figure}
\includegraphics[scale=0.25]{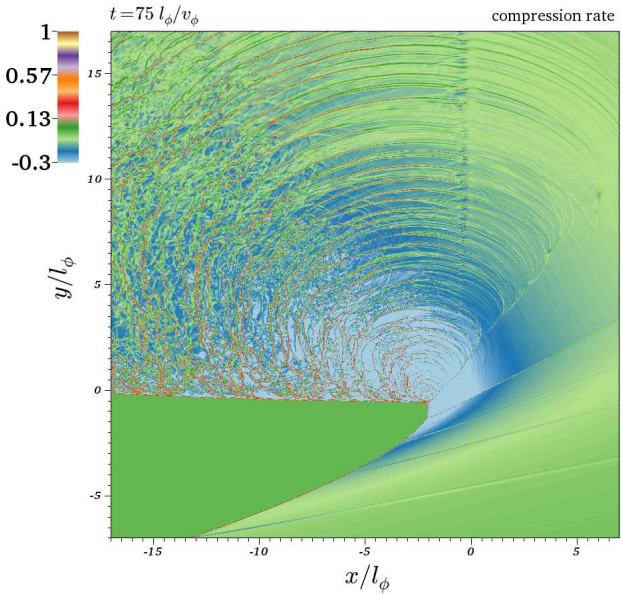}
\caption{Rate of compression ($-\nabla \cdot {\mathbf v}$) in the final, stationary state of the fiducial run.}
\label{Fig:compression}
\end{figure}

The compression rate in the fiducial run is depicted in Figure \ref{Fig:compression} at a late time ($t = 75\, t_\varphi$) for which the flow has entered its final, stationary state; the color scale is chosen to highlight shocks while also making visible large-amplitude sound waves.    
The form of the shock locus corresponds well to our expectations, as depicted in Figure \ref{Fig:ShockFit}.   In particular, the shock curves toward the surface in a manner similar to the  \citeauthor{1964PThPh..32..207I} theory, experiences a kink at finite depth, and is nearly vertical from there to the breakout point.   We measure a terminal shock angle of approximately $36^\circ$ from the vertical, rather than $27^\circ$, just below the kink.

\begin{figure}
\includegraphics[scale=0.25]{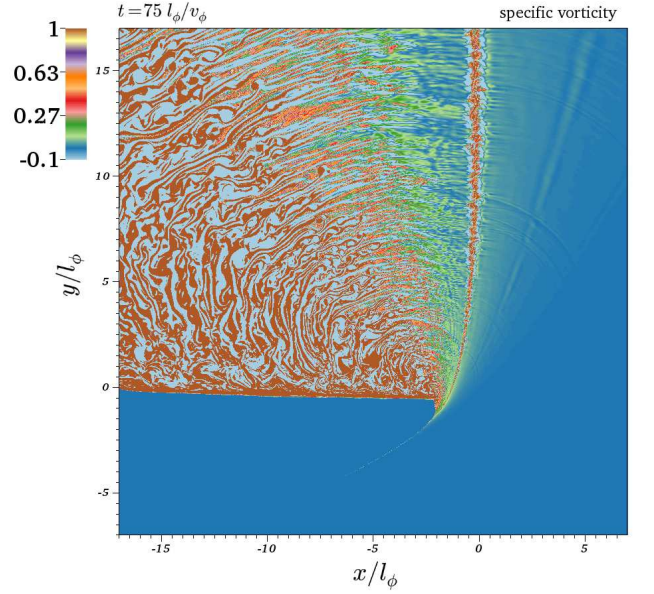}
\caption{Specific vorticity ($[\nabla \times {\mathbf v}]/\rho$) in the final, stationary state of the fiducial run. } 
\label{Fig:vorticity}
\end{figure}

We see in Figure \ref{Fig:compression} that the kink in the primary shock radiates a weak shock downstream, and in Figures \ref{Fig:vorticity} and \ref{Fig:entropy} we see that a vortex sheet and mild entropy discontinuity also emanate from this point, before traveling downstream at a terminal angle $\alpha_f \simeq 90^\circ$.   All of these features are required at a kink in the shock \citep[][\S~102, fig. 83]{1959flme.book.....L}.  A close examination of the highest-resolution subgrid, shown in Figure \ref{Fig:Finest_resolution}, reveals that Kelvin-Helmholz rolls develop along this vortex sheet. 

Another weak shock emerges from the point of breakout, which we take to be the topmost location at which the primary shock meets matter of zero entropy,  $(x,y) = (-2.18,-0.59)\Lphi$; the distribution of the entropy-related quantity $s=P/\rho^\gamma$ is plotted in Figure \ref{Fig:entropy}. 

\begin{figure}
\includegraphics[scale=0.25]{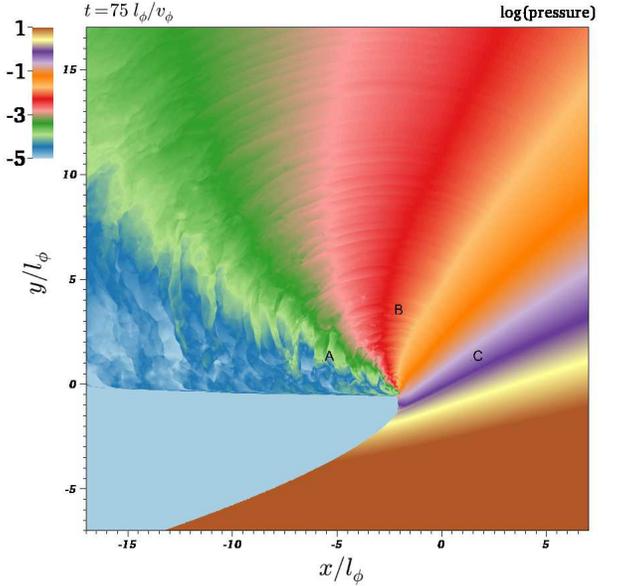}
\caption{Pressure in the final, stationary state of the fiducial run.  The points A, B, and C at which we trace the time histories of various quantities, are overplotted.}  
\label{Fig:pressure}
\end{figure}

\begin{figure}
\includegraphics[scale=0.25]{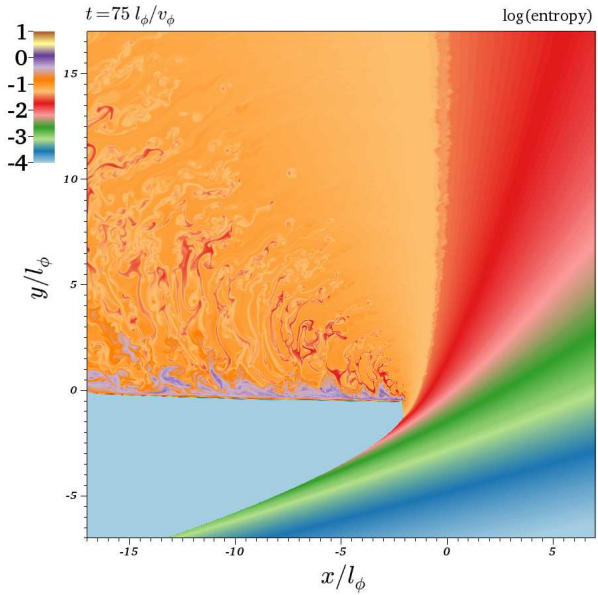}
\caption{The entropy-related quantity $s=P/\rho^\gamma$ in the final, stationary state of the fiducial run; see Figure \ref{Fig:Finest_resolution} for details around the breakout region.}  
\label{Fig:entropy}
\end{figure}

\begin{figure*}[ht]
\includegraphics[scale=0.45]{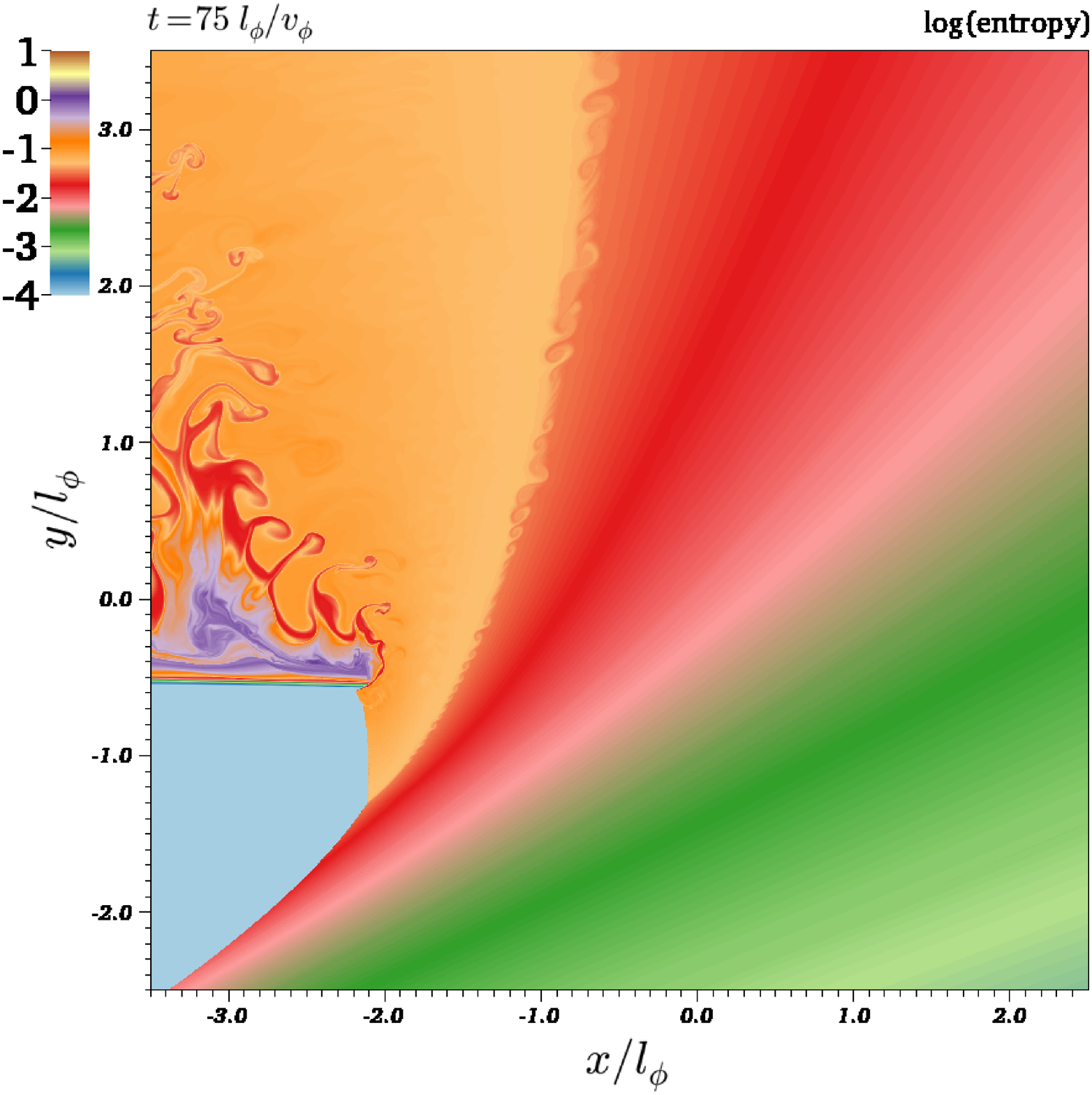}
\caption{The highest-resolution subgrid: entropy $s$ in the final state ($t=75 \Lphi/\vphi$) of the fiducial run.  The compression of the stellar surface upstream of the main shock is easily visible, as is the Kelvin-Helmholz instability which develops along the vortex sheet which emerges from the shock kink.  Note: the image resolution does not approach the native  $4096^2$ resolution of this subgrid. }
\label{Fig:Finest_resolution}
\end{figure*}

Several additional shock  features are apparent in Figure \ref{Fig:compression}, including two very weak shocks emanating from the curving primary shock, and a third weak shock which emanates from the point at which the primary shock meets the lower grid boundary.   We suspect that all of these are related to the effect of finite box size, and are caused by discrepancies between the self-similar boundary conditions and the dynamics of smooth two-dimensional flow.  We examine this question again in \S \ref{SS:Boxsize}, where we vary $\Boxsize$ to assess the influence of such discrepancies.    The secondary shocks are hardly visible in the pressure distribution (Figure \ref{Fig:pressure}), indicating they are much weaker than the primary shock.

\subsection{Distribution of ejecta} \label{SS:ejecta_distribution} 

Observational implications of oblique shock breakout depend critically on the distribution of fluid quantities in the ejecta: that is, the distribution of mass flux and entropy relative to angle in the far field.   A related question is how the final angle of ejection ($\alpha_f$, if measured in the comoving frame) relates to the initial depth ($|y_0|/\Lphi$).   

To gauge these distributions we must assign an approximate $\alpha_f$ to each streamline, as follows.   First we obtain an averaged final stationary state by time-averaging the conserved quantities over the period $40<t/t_\varphi<70$ in the fiducial run.  The velocity vector is not perfectly aligned with $\alpha_f$, thanks to acceleration outside the simulation volume.   However, we can correct for these extra accelerations by appealing to the behavior of the planar, self-similar flow: a mass element's velocity $v'(m)$, expressed in the star's rest frame, is related to its final value $v_f'(m)$ and the sound speed $c_s(m)$ by a definite relationship.  To an accuracy better than 0.2\%, 
\begin{equation}\label{Velocity-corretion}
v_f'(m)- v'(m) = {c_s(m) \over \left[9.1  + \frac{1}{192} \left(v'(m)\over c_s(m)\right)^{1.895}\right]^{1/1.895}}. 
\end{equation} 
Although this strictly applies only to the planar limit which holds deep within the star, we apply the corresponding velocity correction at every point in our solutions. We do this first by evaluating the local stellar-frame Mach number ($v'/c_s$), then computing $v_f'-v'$, then boosting the fluid velocity by this amount in the direction opposite to the local pressure gradient.    We take the terminal angle $\alpha_f$ to be the angle of the resulting velocity. 

This procedure should be valid in the very deep flow, where it captures the behavior of the self-similar solution, and in the shallow, highly supersonic region, where the velocity correction tends to zero.  Any error incurred on intermediate streamlines should decline for large values of the box size parameter $\Boxsize$, because simulations with larger $\Boxsize$ capture regions of higher Mach number.  Fortunately, the angular correction in our fiducial run is at most a few degrees, and our estimate for $\alpha_f$ hardly varies along each streamline.   

\begin{figure}
\includegraphics[scale=0.4]{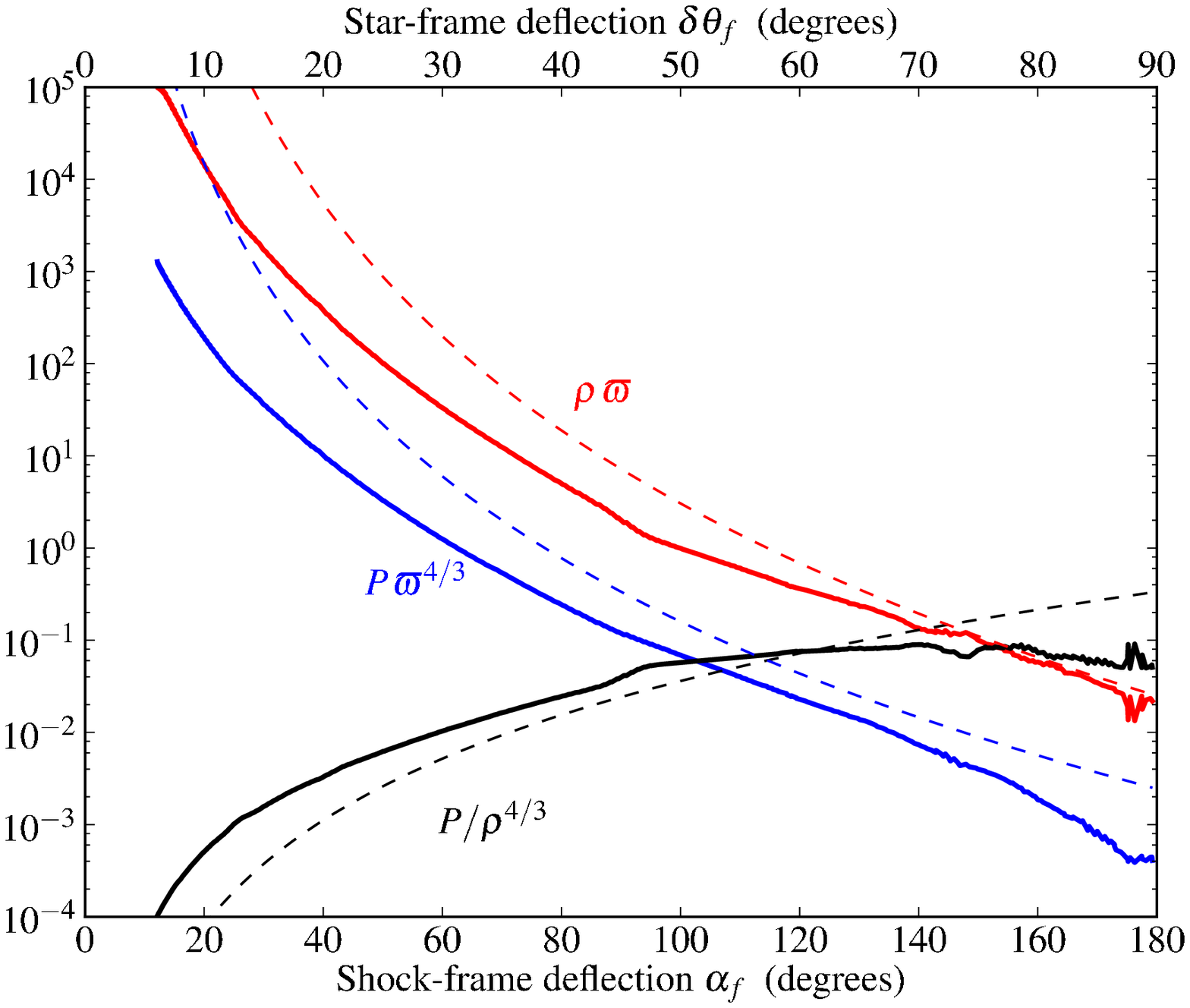}
\caption{Distributions of flow quantities as functions of the estimated terminal angle $\alpha_f$ in the time-averaged final state of our fiducial run.  The quantity $\varpi$ is the distance to the breakout point, so all three quantities plotted are constant along streamlines at large $\varpi$.  All values are normalized to natural units: $\vphi = \rho_\varphi = \Lphi= 1$. Dashed curves represent the known planar, self-similar solution, extrapolated as a power law in $\alpha_f$ from deep planar limit ($\alpha_f\ll1$), as derived in the Appendix. The feature at $\alpha_f = 90^\circ$ corresponds to the shear layer visible in Figure \ref{Fig:vorticity}, and emanates from the kink in the main shock.  }  
\label{Fig:Angular_Distribution}
\end{figure}

The distribution of final quantities is plotted against angle in Figure \ref{Fig:Angular_Distribution}.  We list fits to several angular ranges within the fiducial run in Table \ref{Table:EjectaDistribution}, where we compare to the limit of deep planar flow (derived in the Appendix).  This comparison, indicated by the dashed curves in Figure  \ref{Fig:Angular_Distribution}, indicates that our numerical estimate for $\alpha_f$ is not especially accurate in the deep flow. 

\begin{deluxetable*}{cccccc}
\tablewidth{0pt}
\tablecaption{Ejecta distribution: fiducial run\tablenotemark{a}}
\tablehead{
\colhead{Quantity} & 
\colhead{Fit: all $\alpha_f$ } &
\colhead{ $\alpha_f > 45^\circ$} & 
\colhead{ $\alpha_f > 90^\circ$} & 
\colhead{ Analytical limit ($\alpha_f\rightarrow0$) } & 
\colhead{Equation} 
}
\startdata
$|y_0| /  \Lphi$ &$2.06 \alpha_f^{-1.34} $& $2.59 \alpha_f^{-1.69} $&$ 4.11 \alpha_f^{-2.21}$ & $3.58 \alpha_f^{-1.80}$ & (\ref{eq:deep_limit_y0})\\   
${\rho \varpi / (\rho_\varphi \Lphi)}$ & $28.2 \alpha_f^{-5.82} $&$ 42.9\alpha_f ^{-6.50} $& $38.2 \alpha_f^{-6.37} $&$ 295 \alpha_f^{-8.18}  $& (\ref{eq:deep_limit_rho})\\
${P \varpi^{4/3}/ ( \rho_\varphi \vphi^2 \Lphi^{4/3}) }$&$0.84 \alpha_f^{-5.22} $&$ 2.11 \alpha_f^{-6.64} $& $8.24\alpha_f^{-8.20} $& $8.46 \alpha_f^{-7.11} $ &(\ref{eq:deep_limit_s})\\
${s/(\vphi^2/\rho_\varphi^{\gamma-1} )}$& $0.00807 \alpha_f^{2.61}$ & $0.0122\alpha_f^{1.98}$ & $ 0.053 \alpha_f^{0.31}$ &$ 0.00432\alpha_f^{3.79} $ &(\ref{eq:deep_limit_P}) \\
${\Diffusion / ( \kappa \rho_\varphi \Lphi \vphi/c)} $&$ 126\alpha_f^{-5.96}$ & $169\alpha_f^{-6.36}$ & $388\alpha_f^{-7.32}$ & $124\alpha_f^{-7.17}$ & (\ref{eq:deep_limit_Diffusion})
\enddata
\tablenotetext{a}{Quantities measured in the time-averaged final steady state of the fiducial run at a distance $(8.9\pm0.1)\Lphi$ from the breakout point.}
\label{Table:EjectaDistribution}
\end{deluxetable*}

As viewed in the shock's frame, the stellar envelope moves azimuthally (horizontally) through the shock front and is deflected radially (vertically) by an angle $\alpha$ (see Figure \ref{Fig:ShockFit}), the terminal value of which is $\alpha_f$.   Viewed in the star's frame, matter at rest is struck by an inclined shock and ejected from the star at a terminal angle $\delta \theta_f$ from the radial (vertical) direction.  These two angles are related by the fact that the terminal speed matches the inflow speed in the comoving frame  (a consequence of energy conservation), i.e. $v_f = \vphi$. This implies 
\begin{equation} \label{eq:alphaf_thetaf_relation}
\delta \theta_f = \alpha_f/2. 
\end{equation} 
The outflow speed in the star's rest frame is 
\begin{equation}\label{eq:v_star_frame}
v_f' = \left[2(1-\cos\alpha_f)\right]^{1/2}\vphi = 2\vphi |\sin \delta\theta_f|.  
\end{equation}

\subsection{Unsteady behavior} \label{SS:oscillations} 

\begin{figure}
\includegraphics[scale=0.4]{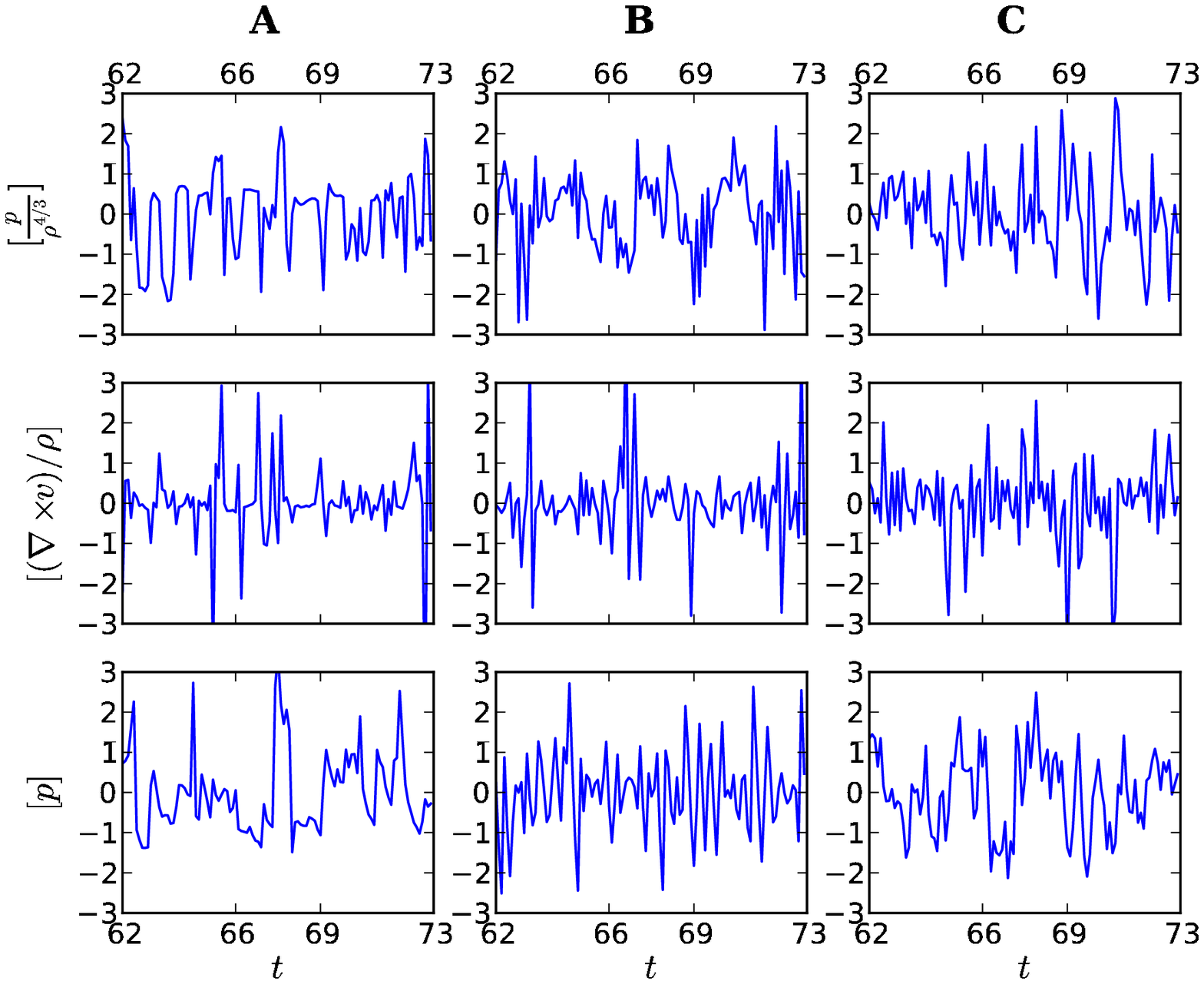}
\caption{Time dependence of entropy ($p/\rho^{4/3}$), specific vorticity ($(\nabla\times v)/\rho$) and pressure ($p$) at the points A, B, and C labeled in Figure \ref{Fig:pressure}.  In each case we plot the mean-subtracted quantity normalized to unit standard deviation over the time interval shown: $[Q] = (Q(t) - \left<Q(t)\right>)/$std$(Q(t))$.  For clarity we show only $62<t/t_\varphi<73$.} 
\label{Fig:Time_dependence}
\end{figure}

Our final state is not steady, as it exhibits acoustic oscillations, entropy fluctuations, and vortices.  Unsteadiness is evident in  the time history of pressure, entropy, and specific vorticity at the locations we plot in Figure \ref{Fig:Time_dependence}.   Our box-size and resolution studies show that the dominant oscillation frequencies are independent of $\Boxsize$ and only weakly dependent on $\Resolution$, so we infer that they are representative of the physical solution.   There is no sign of any high-pitched acoustic oscillations, those of periods $\leq(10 t_\varphi/\Resolution = 10\,\Delta x_{\rm min}/\vphi)$, which would be expected from interactions between the flow and the discretized grid. 

The oscillation periods in vortical, entropy, and pressure variations are similar, and range from a fraction of $t_\varphi$ to several $t_\varphi$.  
Because the shock structures are quite steady, we interpret the entropy fluctuations as being due to vortices moving fluid across the time-averaged streamlines.     Unlike vorticity and entropy, pressure oscillations are restricted to certain flow zones: in particular, they are confined to be downstream of the shock  which emanates from the breakout point.   A plausible source for all these oscillations is the unstable break-up of the shocked slab discussed below. 

\subsection{Interaction of ejecta with the stellar surface} \label{SS:ejecta_surface_interaction} 

Paper 1 raised the possibility that interaction between the ejecta and the stellar atmosphere might be the reason  that flow around the breakout point cannot be described with a steady, self-similar solution.    In this case the pressure of the ejecta, enhanced perhaps by shocks in the region where it interacts with the star, would drive a weak shock and a downward flow in the upstream (otherwise undisturbed) stellar envelope.   Self-similarity  is then destroyed, or at least dramatically changed, because the density no longer varies as the $n$th power of distance from the breakout point. 

This scenario plays out precisely as described within our simulations.   Pressure in the surface-skimming ejecta ($\alpha_f \simeq \pi$) compresses the outermost regions of the stellar envelope, depressing its interface with the stellar surface from $y=0$ to $y=-0.59\Lphi$ in the fiducial run.  This effect is most visible in Figure \ref{Fig:ShockFit}.  The primary shock front shows a visible feature where it intersects this layer, and we strongly suspect that some of the oscillations and vortices in the downstream flow are the consequence of this layer's evolution behind the shock.
In Figure \ref{Fig:compression}, the local compression rate shows finite-amplitude sound waves or weak shocks in those ejecta that skim the stellar surface, and these radiate as sound waves into adjacent streamlines.  

Let us compare the depression of the stellar surface layer with a simple calculation based on the observed ejecta pressure, which in the fiducial run is $P(\alpha_f = \pi) \simeq 10^{-3.5} \rho_\varphi \vphi^2 (\Lphi/\varpi)^{4/3}$.    Except for a minor correction due to the acceleration of the shell, this matches the pressure $(6/7)  \dot y_{\rm as}^2\rho_0(y_{\rm as})$ behind the atmosphere shock (subscript $as$), which we idealize as moving vertically downward.  The instantaneous speed of the envelope shock, $\dot y_{as}$,  equals $-\vphi\, dy_{\rm as}/d\varpi$ because of the motion of the breakout point relative to the stellar surface.   Using $\rho_0(y) =(-y/\Lphi)^n  \rho_\varphi $, this implies $dy_{\rm as}/d\varpi \simeq -0.019 (\Lphi/|y|)^{n/2} (\Lphi/\varpi)^{2/3}$.   Setting $n=3$ and integrating from $y_{\rm as}=0$ at $\varpi = \varpi_0$ (as in the upstream boundary of our simulations), we find 
\begin{equation}\label{eq:shock-depression} 
y_{\rm as} \simeq - 0.46 \Lphi \left[ \left(\varpi_0\over \Lphi\right)^{1/3} - \left(\varpi \over \Lphi\right)^{1/3}\right]^{2/5}.  
\end{equation} 
As depicted in Figure \ref{Fig:ShockFit}, this is an excellent approximation to the shape of the inward shock in the fiducial run ($\varpi_0=14.2$); it predicts this shock will be found at a depth of $y_{\rm as}(\varpi=0) \simeq -0.65 \Lphi$ near the breakout point.  Furthermore, given the compression factor of 7, it predicts that the effective breakout point will be at six-sevenths of this depth, or $y=-0.56 \Lphi$.  This is very close to the measured value of $-0.59\Lphi$ we list in Table \ref{Table:Runs}.  
  
It is important to note that the depression of the stellar surface diverges, albeit very slowly, with distance from the breakout point.  If we associate the initial distance $\varpi_0$ with the stellar radius, equation (\ref{eq:shock-depression}) indicates that the breakout depth will be about $0.4 (R_*/\Lphi)^{2/15} \Lphi$.   This raises the possibility that the flow on scales of $\Lphi$ will be qualitatively changed if $\Lphi\ll R_*$ relative to what we can simulate in a finite box. We consider this and other  finite-volume effects in \S \ref{SS:Boxsize}.

\section{Numerical effects } \label{S:NumericalEffects}
To disentangle physical phenomena from those imposed by the numerical implementation,  we independently vary both the physical resolution $\Resolution$ and the size $\Boxsize$ of the simulation volume.  The simulations we use for this are listed in Table \ref{Table:Runs}.   In the subsections below we concentrate on these parameters' effects, some of which we have already mentioned.

\begin{deluxetable*}{lccccccccc}
\tablewidth{0pt}
\tablecaption{Simulations used to explore resolution and box size effects}
\tablehead{
\colhead{Run} & 
\colhead{$\Resolution$ } &
\colhead{$\Boxsize$} & 
\colhead{$\frac{x}{x_\varphi}$ range} & 
\colhead{$\frac{y}{y_\varphi}$ range} & 
\colhead{$\max \frac{t}{t_\varphi}$ } & 
\colhead{Levels }  & 
\colhead{Breakout point\tablenotemark{a} }
 & 
\colhead{Kink altitude\tablenotemark{b}} & 
\colhead{Match altitude\tablenotemark{c}}
}
\startdata
Fiducial 	& 683 & 24 & -17 to 7 & -7 to 17 & 80 & 3 & $(-2.18, -0.59) \Lphi$  & -1.32$\Lphi$  &-6.48$\Lphi$\\   
R341 	& 341 & 24 & -17 to 7 & -7 to 17 & 80 & 3 & $(-2.16, -0.53) \Lphi$ & -1.30$\Lphi$ & -6.4$1\Lphi$ \\
R171	& 171 & 24 & -17 to 7 & -7 to 17 & 73.6 & 3 & $(-2.12, -0.56) \Lphi$ & -1.28$\Lphi$ &  -6.26$\Lphi$ \\
B12 		& 683 & 12 & -8 to 4   & -4 to 8   & 80  & 3 & $(-1.64, -0.51) \Lphi$ & -1.23$\Lphi$ &  -4.60$\Lphi$ 
\enddata
\tablenotetext{a}{Location, in the final steady state, where the main shock breaches undisturbed stellar matter; its $y<0$ because of the downward reaction of the upstream stellar surface to out-flowing ejecta.}
\tablenotetext{b}{Altitude, in the final steady state, of a kink in the shock angle.}
\tablenotetext{c}{Altitude, in the final steady state, of the best-fit matching between the deep asymptotic solution and the \citet{1964PThPh..32..207I} approximate theory, as in Figure \ref{Fig:ShockFit}.
\vspace{0.25in} }
\label{Table:Runs}
\end{deluxetable*}

\subsection{Finite resolution} \label{SS:Resolution} 
\begin{figure}
\includegraphics[scale=0.4]{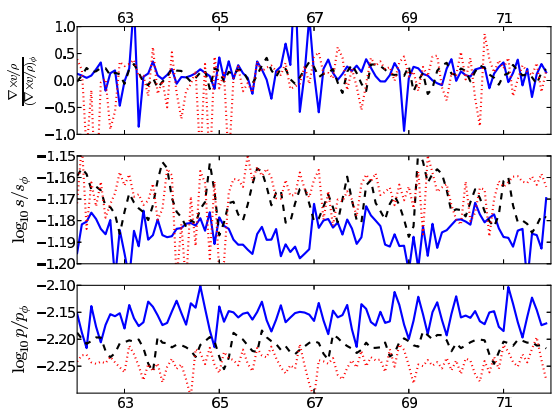}
\caption{Time dependence of specific vorticity ($(\nabla\times v)/\rho$), entropy ($s=p/\rho^{4/3}$), and pressure ($p$) at point B (Figure \ref{Fig:pressure}) for runs with different numerical parameters. In each plot the fiducial run is a solid blue curve; the quarter-resolution run R171 is a black dashed curve; and the half box-size run B12 is a red dotted line.  For clarity we show only $62<t/t_\varphi<73$.}   
\label{Fig:Time_dependence_comparison}
\end{figure}

Our resolution study consists of simulations with identical grid geometries and identical $\Boxsize$, in which $\Resolution$ varies between 683 and 171, corresponding to individual grids (at each resolution level) which vary between $4096^2$ and $1024^2$. Pressure, entropy, and specific vorticity distributions for the lowest and highest resolution runs are depicted in Figure \ref{Fig:Resolution_and_Boxsize_comparison}. Figure \ref{Fig:Time_dependence_comparison} shows the effects of $\Resolution$ on the oscillation modes.

The structure of the shock, the shapes of the post-shock streamlines, and the final entropy distribution are all very insensitive to $\Resolution$.   Oscillation frequencies are not strongly $\Resolution$-dependent, except that the lower-resolution runs show a suppression of some of the short-wavelength motions.   It appears  that there is very little difference between the  highest-resolution runs; suggesting that these faithfully represent the continuum limit $\Resolution\rightarrow \infty$.  However it is always possible for very slow $\Resolution$ dependence, or phenomena which depend on threshold Reynolds numbers, to spoil this fidelity. 

We have not conducted any runs of sufficiently low resolution, $\Resolution\lesssim1$, such that the region of non-radial flow is poorly resolved.  However we speculate in \S \ref{S:conclusions} about what our results might imply for global simulations which lack resolution of the surface layers. 

\subsection{Finite volume} \label{SS:Boxsize} 
Our simulation volume cannot approximate the ideal infinite case, especially as we were unable to use more than three levels of refinement.  What makes this limitation especially severe for our physical problem is the fact that our boundary conditions are derived from the planar limit, which is only valid for matter which originates deep within the star (streamlines with $y_0\ll -\Lphi$).  Because the shock locus curves rapidly upstream ($x_s \propto |y_s|^{1+\lambda}$), it is difficult to reach these deep streamlines in a simulation of finite volume which must also capture the obliquity zone and the outflow region.   We therefore expect to see $\Boxsize$-dependent features associated with a readjustment from the flow we impose at the boundaries toward something more representative of the ideal $\Boxsize\rightarrow\infty$ solution.   

Prime candidates for these features are the weak shocks which emanate from the lower grid boundary: one at $x=0$, and another where it meets the primary shock.   A couple weak shocks are also launched from the curving primary shock, where there is no clear interaction with the stellar surface, and these may also depend on the simulation volume.  As for the geometrical shape of the primary shock, the approximate theory of \citet{1964PThPh..32..207I} would suggest that this is quite insensitive, in a power-law atmosphere, to the details of the simulation -- at least, up to an overall scaling (the value of $\Lphi$) and the absolute location of shock breakout ($x_s$ for $y_s=0$).  

\begin{figure*}[ht]
\centering
\subfigure[Pressure $P$, fiducial run]{\includegraphics[scale=0.15]{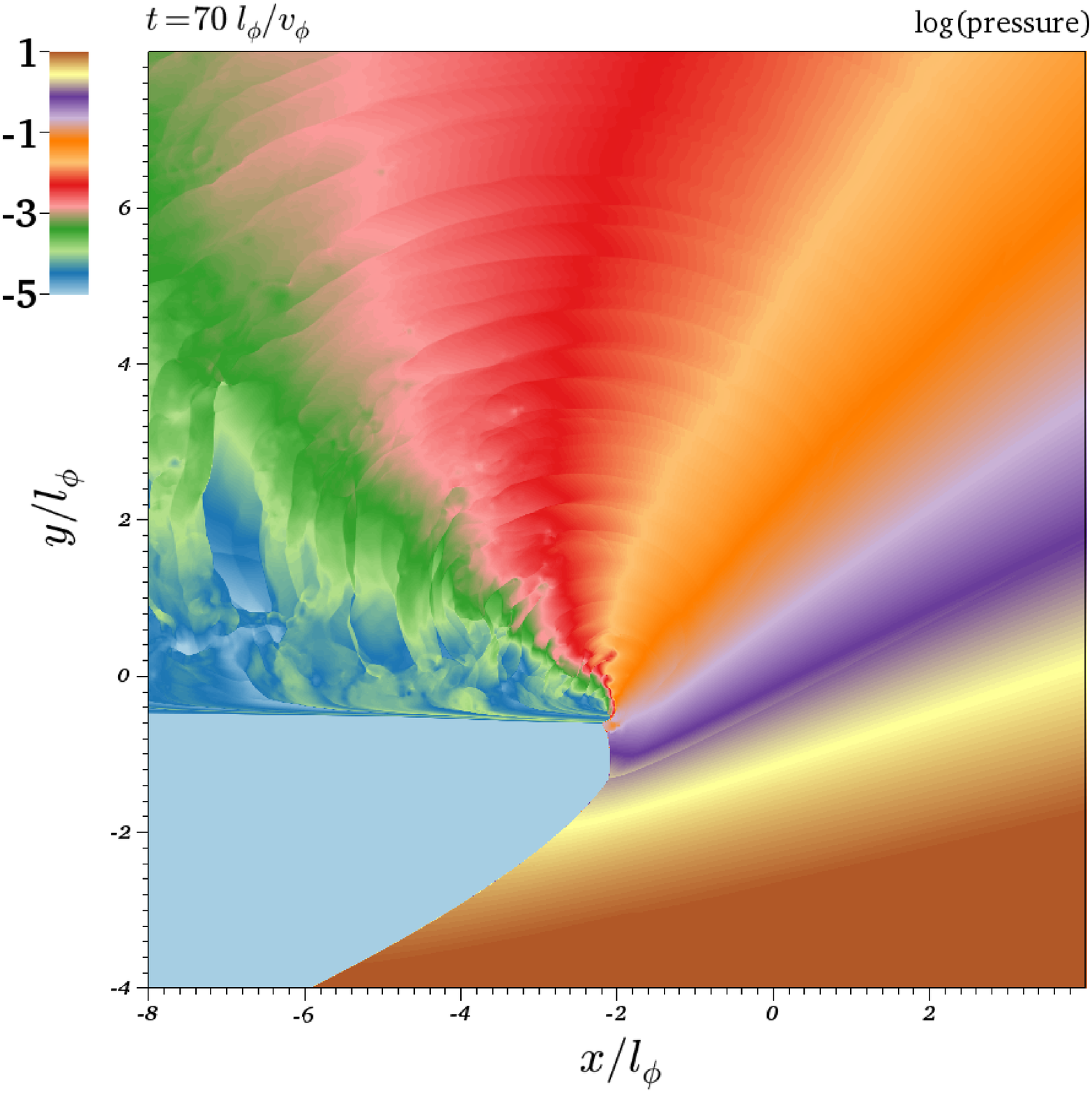}} 
\subfigure[Entropy $s$, fiducial run]{\includegraphics[scale=0.15]{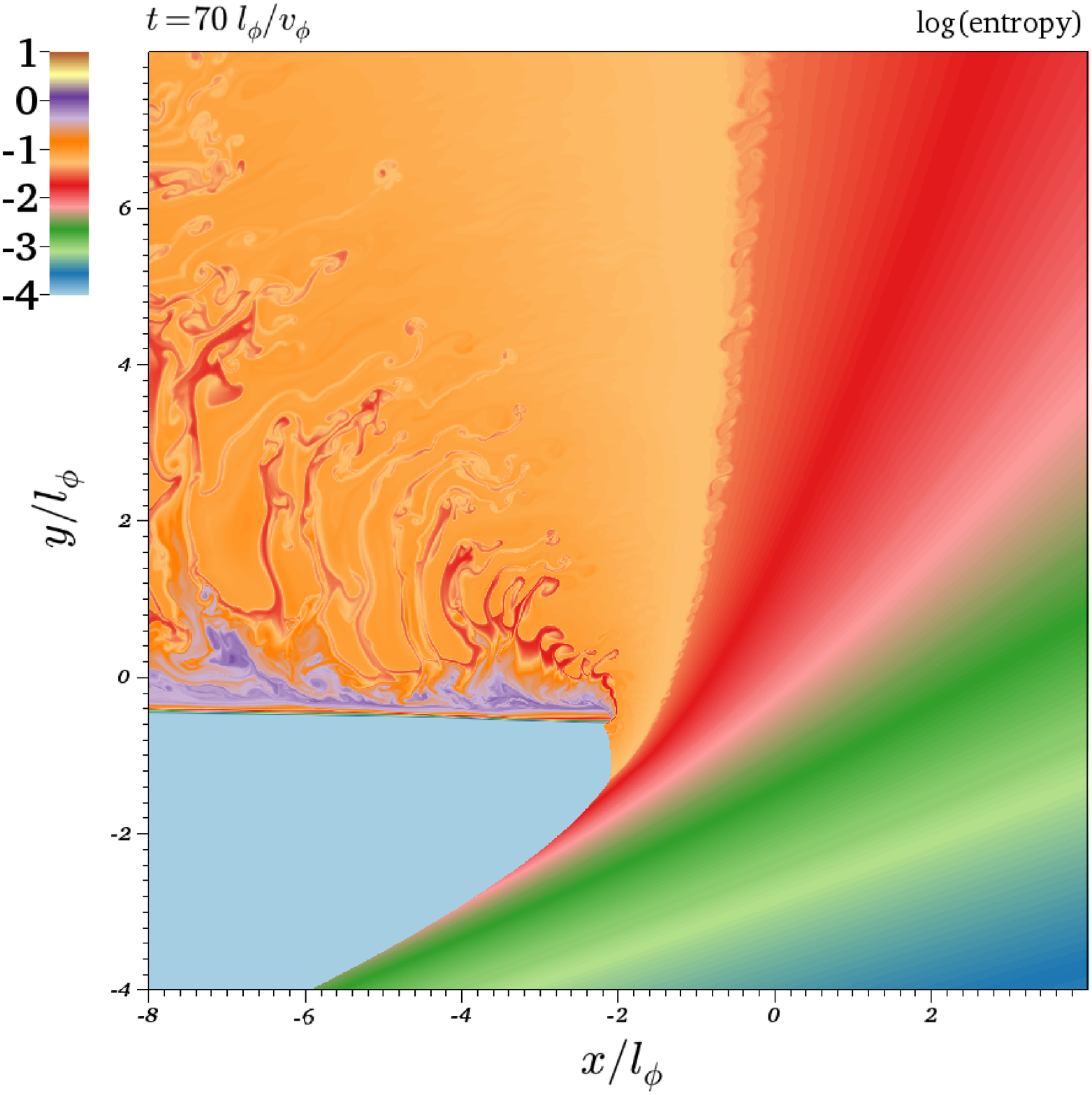}}
\subfigure[Specific vorticity, fiducial run]{
\includegraphics[scale=0.15]{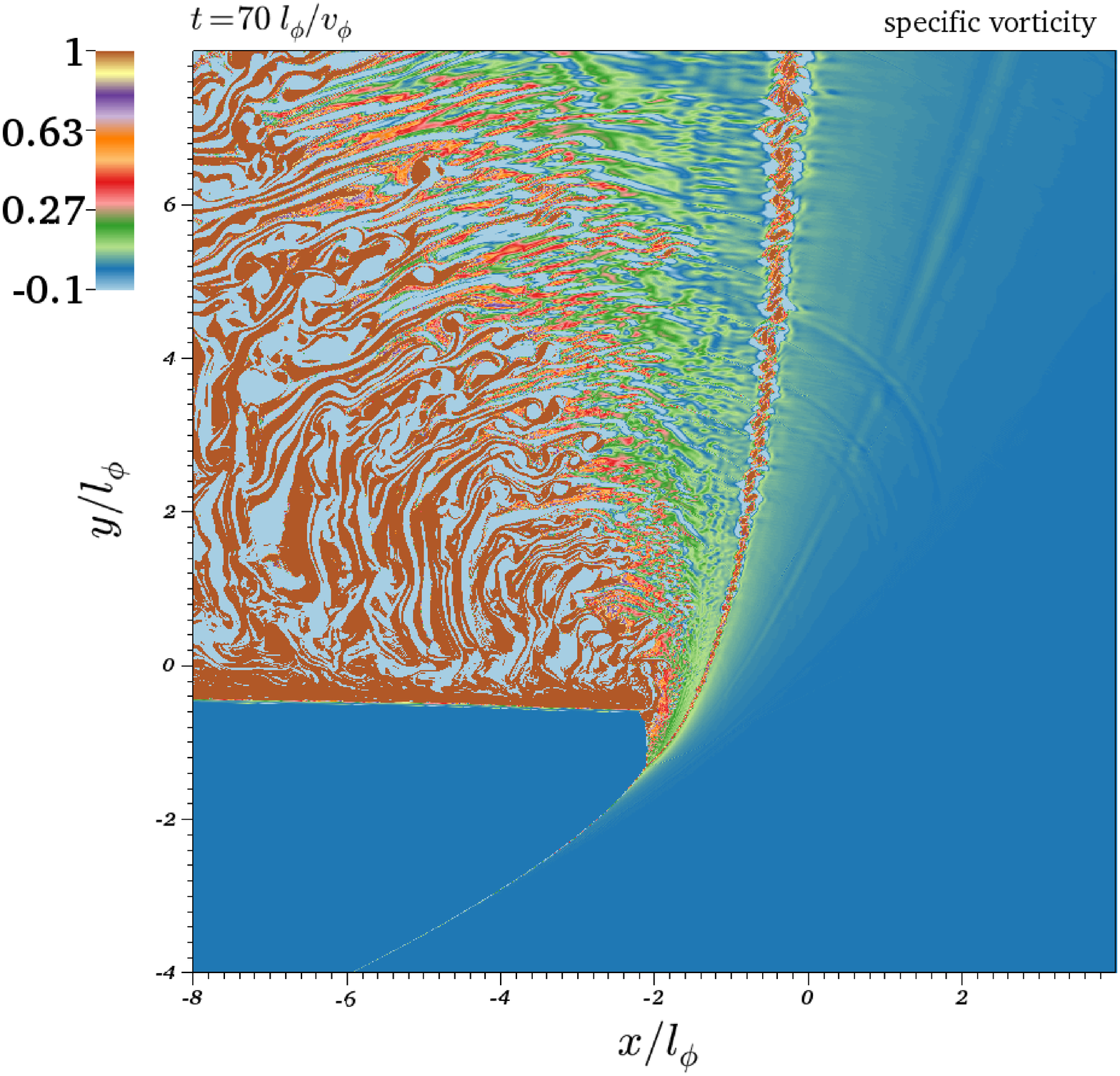}}

\subfigure[$P$, quarter-resolution run R171]{ \includegraphics[scale=0.15]{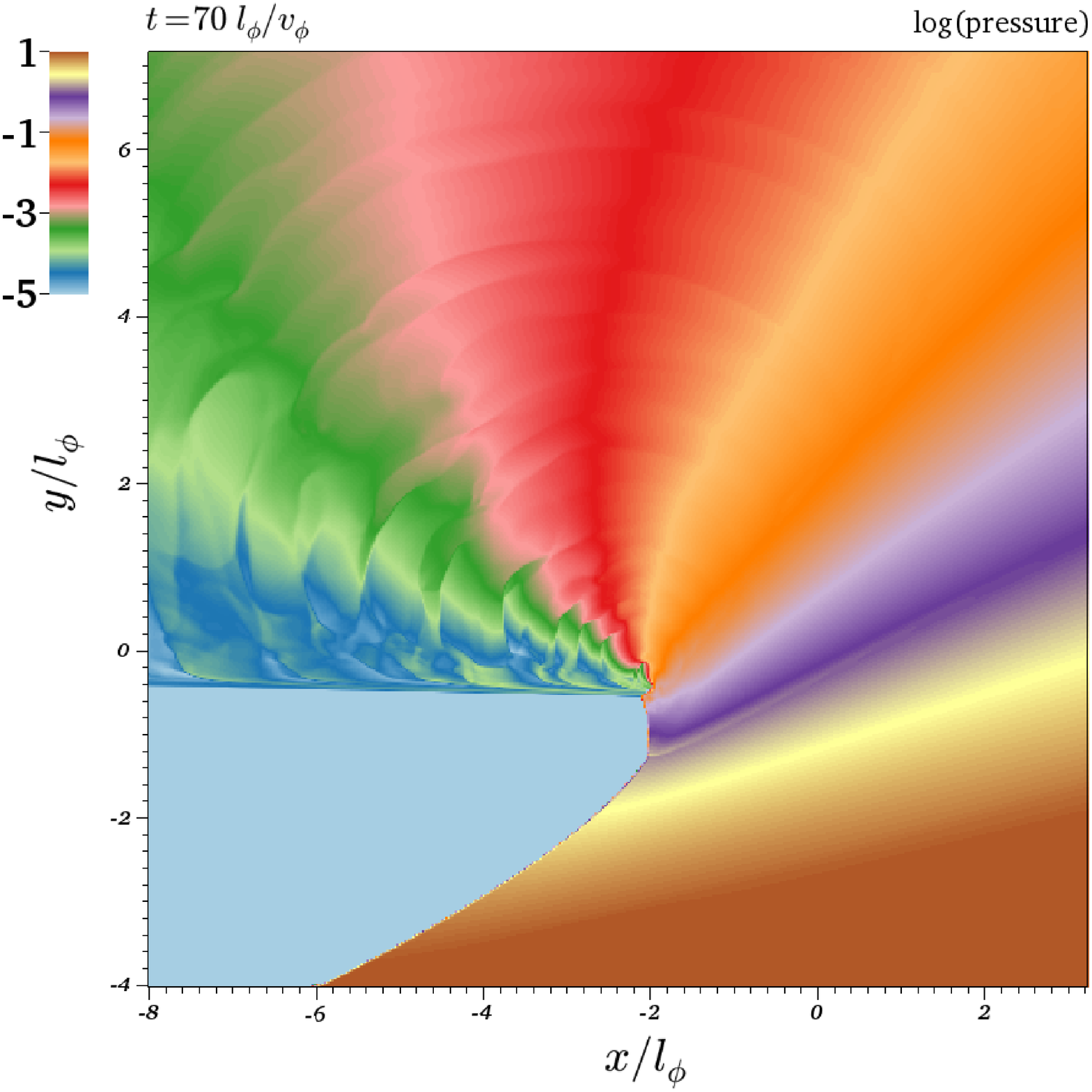} } 
\subfigure[$s$, quarter-resolution run R171]{\includegraphics[scale=0.15]{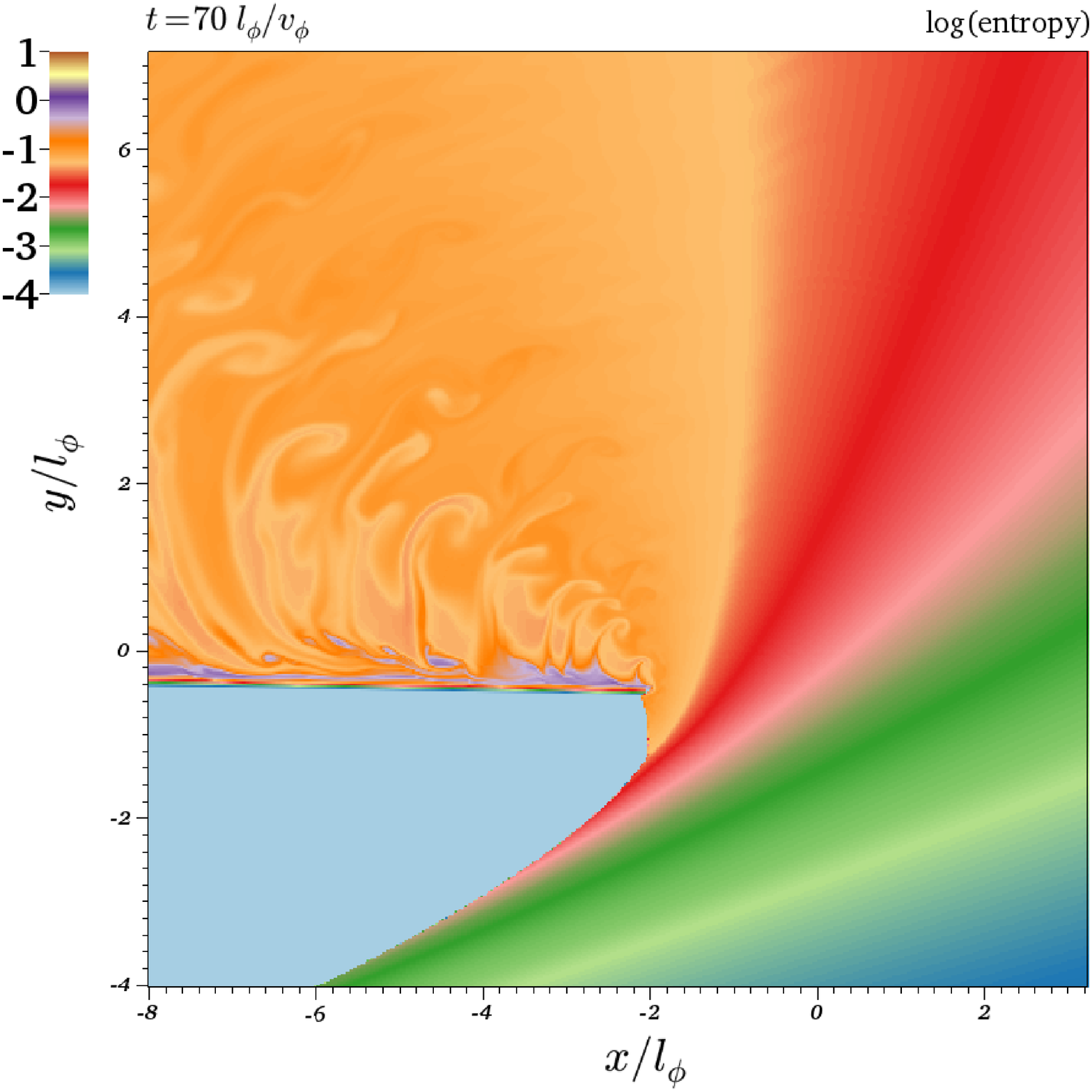}}
\subfigure[Specific vorticity, run R171]{\includegraphics[scale=0.15]{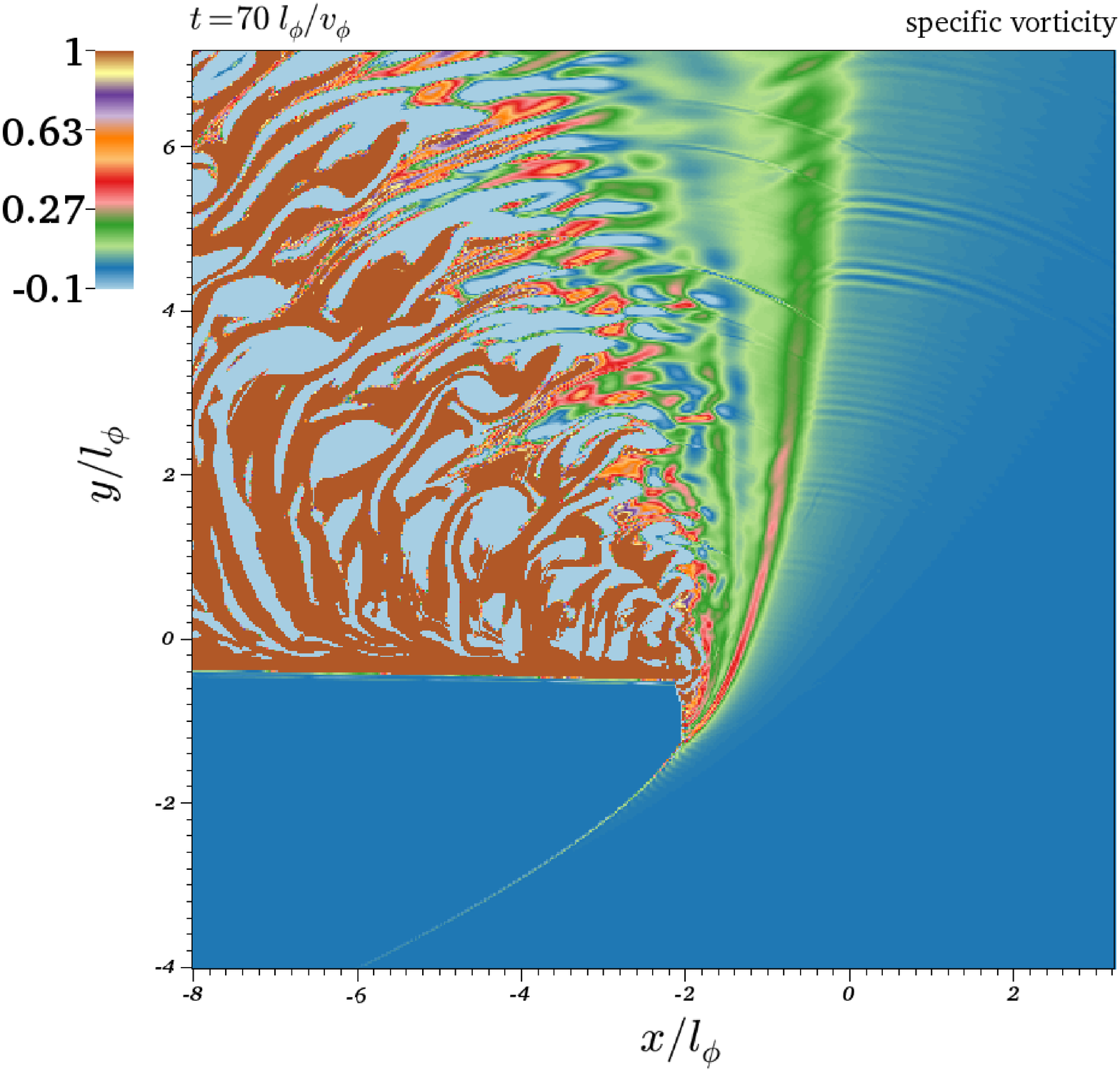}}

\subfigure[$P$, half-box-size run B12]{ \includegraphics[scale=0.15]{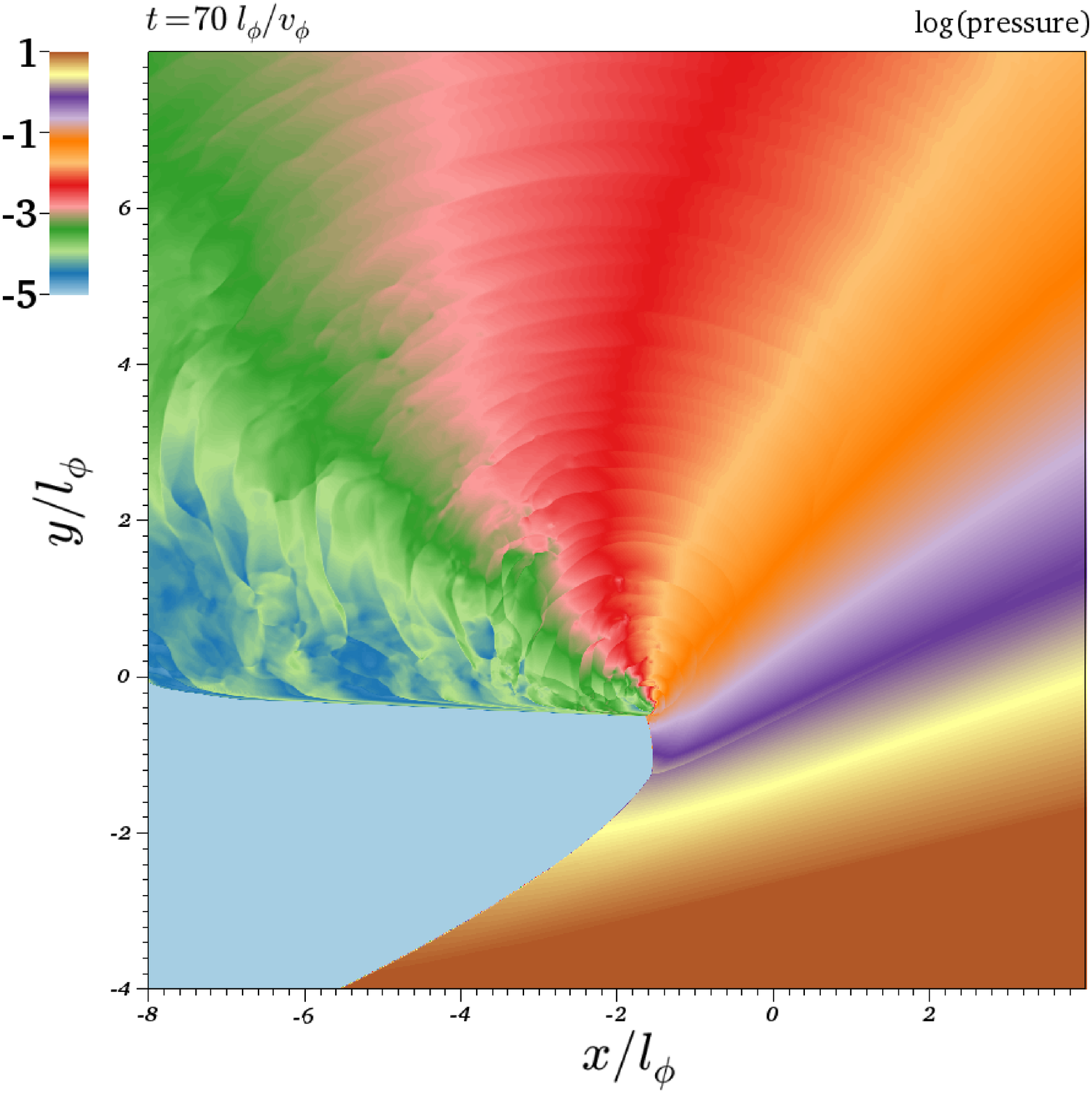} } 
\subfigure[$s$, half-box-size run B12]{\includegraphics[scale=0.15]{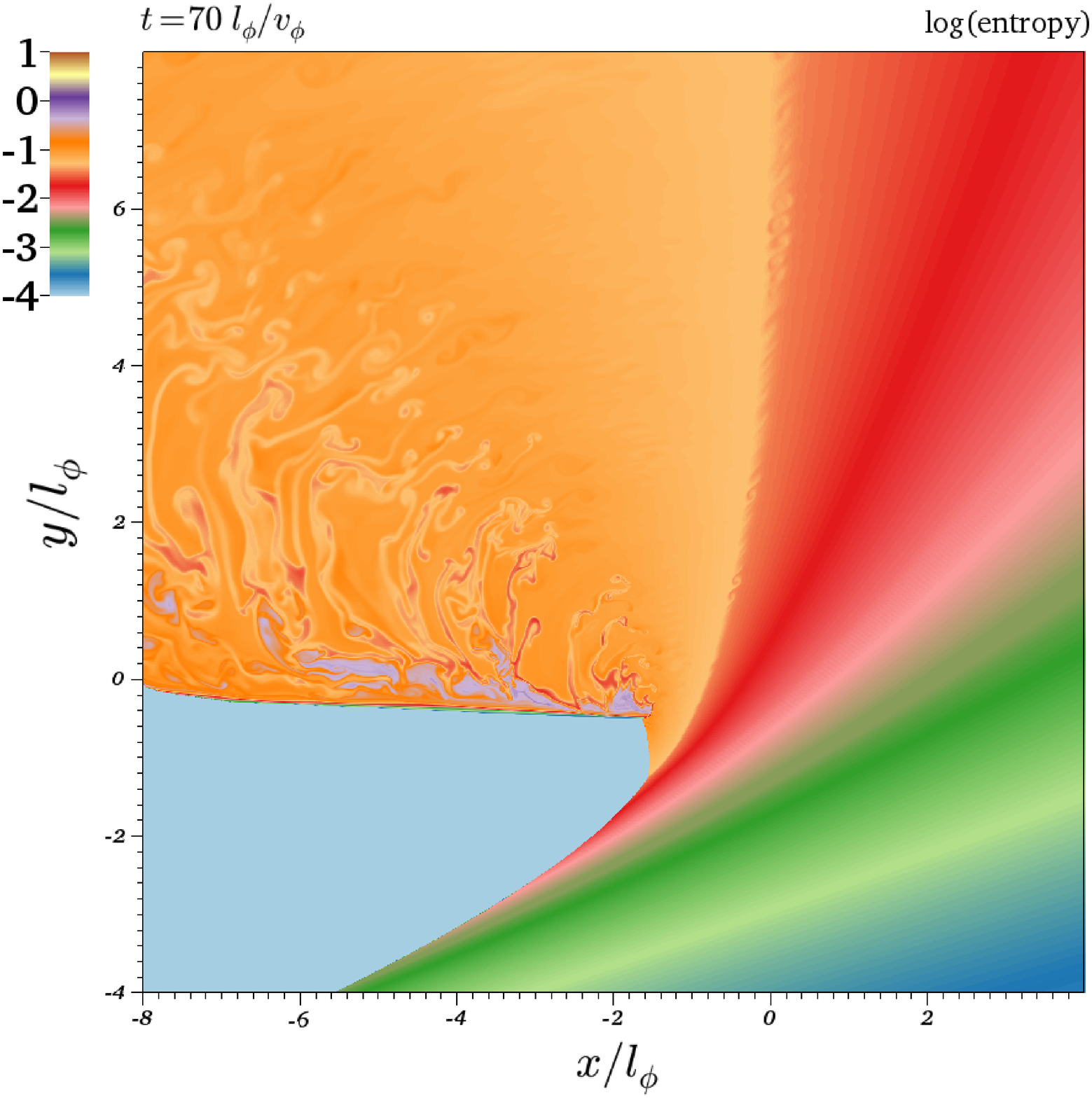}}
\subfigure[Specific voriticity, run B12]{\includegraphics[scale=0.15]{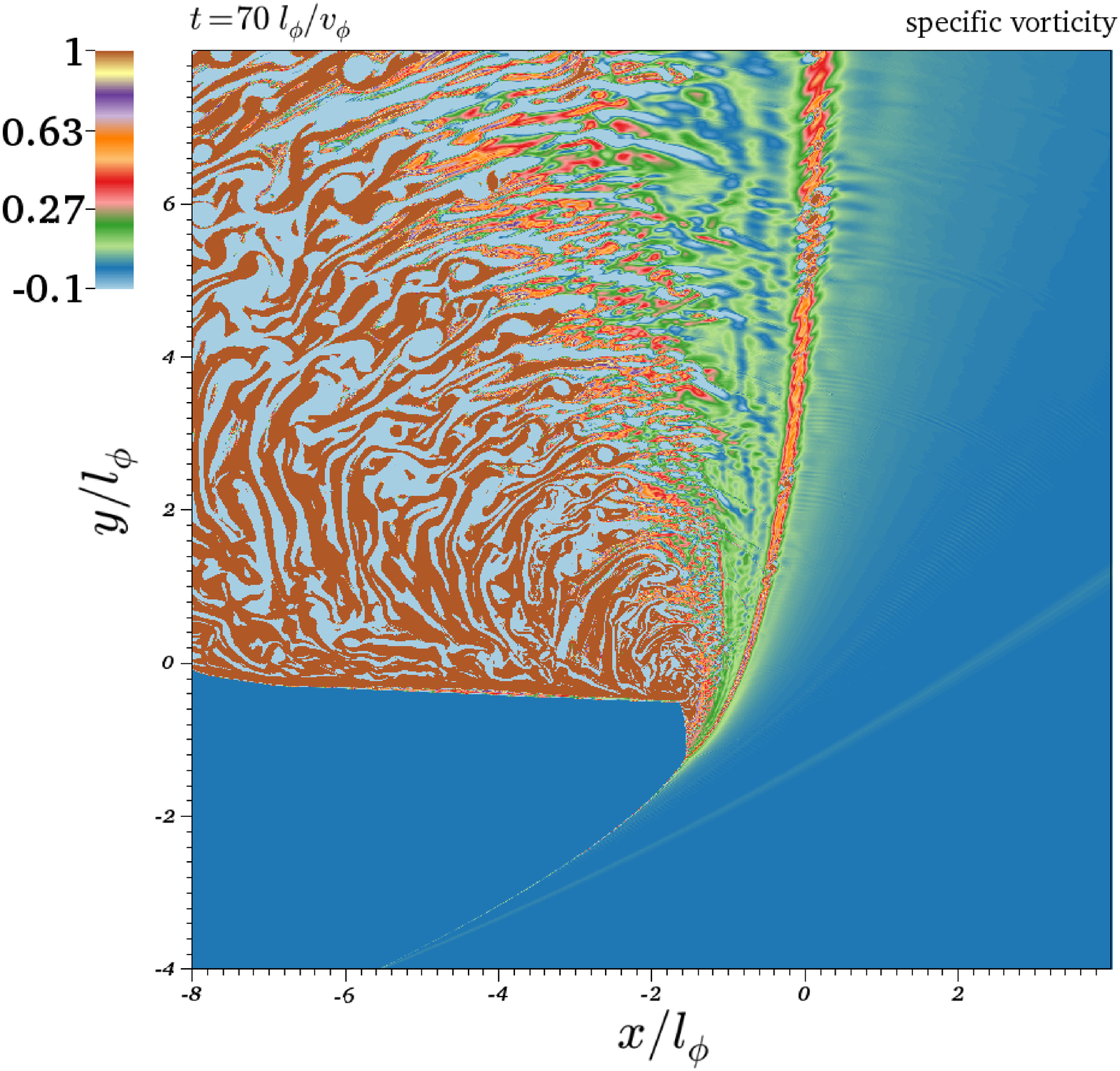}}

\caption{Comparison of the pressure, entropy, and specific vorticity between runs with different numerical parameters.  All snapshots are taken at $t= 70 t_\varphi$, and the region plotted is covered by all three runs.  The effect of lower resolution can be seen in the spatial scale of oscillations in the middle panels, and the effect of a smaller box size can be seen in the breakout location and the appearance of a weak shock discontinuity on the lower panels. }
\label{Fig:Resolution_and_Boxsize_comparison}
\end{figure*}

All of these expectations are validated in the first and third rows of  Figure \ref{Fig:Resolution_and_Boxsize_comparison}, which compare the pressure in simulations of different $\Boxsize$ and identical $\Resolution$.  Their primary shock structure is very similar, except that going from the lower to the higher-resolution run, the depth of the shock kink is greater by $0.04\Lphi$ and the point of breakout is shifted to the left by $0.06\Lphi$.  The shapes of the streamlines are very similar, and they are almost identical once we apply a translation and rescaling to bring the breakout point and shock kink together.  Those features which genuinely differ between the two runs are those which originate from the lower boundary.   The weak shocks radiated by the primary shock also depend weakly on the distance to the box boundary. 

However, as we note in \S \ref{SS:ejecta_surface_interaction}, there is another feature of the flow which clearly depends on its finite $\Boxsize$: the ejecta-envelope interaction upstream of the breakout region.  Our calculation there suggests the breakout point descends as $\Boxsize^{2/15}$ in runs with progressively larger $\Boxsize$.

\section{Radiation Diffusion and Appearance}\label{S:OpticalDepth}

\begin{deluxetable*}{lccccccccc}
\tabletypesize{\scriptsize}
\tablewidth{0pt}
\tablecaption{Characteristics of model core-collapse supernovae\tablenotemark{a}}
\tablehead{
\colhead{Model} & 
\colhead{$\begin{array}{c} M_{\rm ej} \\ (M_\odot)\end{array} $} &
\colhead{$\begin{array}{c} R_*\\ (R_\odot)\end{array} $} & 
\colhead{$n$} & 
\colhead{$\frac{\rho_h R^3}{M_{\rm ej}}$ } & 
\colhead{$\frac{v_* t_{\rm se}}{R_*}$ } & 
\colhead{$\frac{R_*}{\Lphi}$} & 
\colhead{$\frac{c}{2\vphi}$} & 
\colhead{$\alpha_\Diffusion$} 
}
\startdata
RSG\tablenotemark{b}    	& 14 & 490 & 1.20 
    & 0.566  
  &0.520 & $\left({\cal V}_\varphi\over 0.42\right)^{4.3}  $  
  & ${41E_{51}^{-1/2}\over {\cal V}_\varphi}$  
    &$\frac{\pi}2  \left(2.2 E_{51}^{0.06}\over {\cal V}_\varphi \right)^{1.1}$ \\
BSG\tablenotemark{c}   	& 15 & 50   & 3.88  
   & 0.0254 
     & 0.409 &$\left({\cal V}_\varphi \over 0.60\right)^{1.4}    $   
     & $\frac{33E_{51}^{-1/2}}{{\cal V}_\varphi}$ 
   &$\frac{\pi}2  {6.43E_{51}^{0.09}\over {\cal V}_\varphi }$\\
Ic\tablenotemark{d} 		& 5   & 0.2   & 5.12 
  & 0.350    
   & 0.435 &$\left({\cal V}_\varphi\over 0.40\right)^{1.1}    $   
   &$\frac{ 21 E_{51}^{-1/2}}{{\cal V}_\varphi }$  
  &$\frac{\pi}2  {41 E_{51}^{0.09}\over {\cal V}_\varphi }$
\enddata
\tablenotetext{a}{Polytropic parameters $n$ and $\rho_h$ are fit to hydrostatic regions in the outer 20\% of the stellar radius.  Shock parameters $\beta$ and $C_2$ are derived from fits by \citet{2013ApJ...773...79R}; coefficient $C_1$ is adjusted to the stellar profile as described by \citet{2001ApJ...551..946T}.  The \citeauthor{2001ApJ...551..946T} shock velocity model is used to calculate $v_*t_{\rm se}/R_*$.   The symbol ${\cal V}_\varphi$ indicates the ratio $\vphi t_{\rm se}/v_*$; in the bipolar explosion model of Paper 1, ${\cal V}_\varphi = 1/[2 \varepsilon \sin(2\theta)]$ where $\varepsilon$ is an elongation factor. }
\tablenotetext{b}{Model s15s7b2 of \citet{1995ApJS..101..181W}, provided by Stan Woosley.}
\tablenotetext{c}{\citet{1990ApJ...360..242S} model for the progenitor of SN 1987A, provided by Ken'ichi Nomoto.} 
\tablenotetext{d}{ Model CO6 of \citet{1999ApJ...516..788W}, for the progenitor of SN 1998bw, provided by Stan Woosley.  Coefficients listed here are modified slightly by relativistic effects for energies $E_{51}>10$.\\ } 
\label{Table:Examples}
\end{deluxetable*}

In the diffusion approximation the radiative flux ${\mathbf F} = - c (\nabla P_{\rm rad})/(\kappa \rho)$ defines a diffusivity $\nu_{\rm rad} = c/(3\kappa \rho)$ and a  photon diffusion speed $v_{\rm diff} = |{\mathbf F}|/(3 P_{\rm rad}) = \nu_{\rm rad}/{\cal L}_p = c /(3\kappa \rho{\cal L}_P)$, where ${\cal L}_p = P_{\rm rad}/|\nabla P_{\rm rad}|$ is the local scale length of radiation pressure. In a radiation pressure-dominated flow $\nu_{\rm rad}$ applies to the diffusion of pressure.  In the context of a breakout flow, therefore, the dynamical effects of radiation diffusion are  determined by the local P\'eclet number $\Diffusion = v/v_{\rm diff} = {\cal L}_p v/\nu_{\rm rad}$: diffusion is negligible where $\Diffusion>1$ and strong where $\Diffusion <1$.   In the context of real explosions, what does this mean for the validity of our adiabatic simulations and for observations of oblique breakouts? 

Within our flow $\Diffusion$ becomes constant with distance $\varpi$ along each outflow streamline, because ${\cal L}_p \propto \varpi$, while $\rho \propto \varpi^{-1}$ and $v\rightarrow \vphi$.   Fitting the angular dependence over all angles within our  fiducial run we find that $3 \rho {\cal L}_p v \simeq D_1 \, \alpha_f^{-d} \rho_\phi \Lphi \vphi $ with $D_1=126$ and $d=6$  (see Table \ref{Table:EjectaDistribution} and the Appendix  for more information) so that
\begin{equation}\label{eq:Diffusion}
 \Diffusion \simeq {D_1\over \alpha_f^{{\delta_{\Diffusion}} } }\kappa \rho_\varphi \Lphi {\vphi\over c}.  
\end{equation}
For comparison, an extrapolation from the deep, planar flow would give $D_1=122$ and ${\delta_{\Diffusion}} =7.2$.  

In the numerical fit, the prefactor $126\alpha_f^{-6}$ takes a minimum value of 0.13 in the turbulent boundary layer between ejecta and star ($\alpha = \pi$), where $\Diffusion$ fluctuates.  Inspecting this layer within the numerical solution, we see that $\Diffusion$ dips to a minimum of $0.05 \kappa \rho_\varphi \Lphi \vphi/c$ where  $\alpha_f \simeq 174^\circ$. 
Purely adiabatic calculations are appropriate (and breakout emission is entirely suppressed) where $\Diffusion\gg 1$, so our neglect of diffusion along the stellar surface requires $ v_\phi \rho_\varphi \Lphi \kappa /c \gg 20$.   If this condition is not satisfied, the deep radial flow will be unaffected but radiation diffusion will limit the hydrodynamic deflection to those streamlines that have $\Diffusion\gtrsim 1$.  (To compare $\Diffusion$ to the parameter $\vphi/\hat v_{s,{\rm max}}$ used in Paper 1, note that $3\rho_\varphi \Lphi \vphi/c = (n+1) (\vphi/\hat v_{s,{\rm max}})^{-(\gamma_p/\beta-1)}$.  For the $n=3$ case, this implies $\alpha_{\cal D}/(\pi/2) = 1.49 (\vphi/\hat v_{s,{\rm max}})^{-1.03}$.) 

A similar criterion, discussed in Paper 1, compares the characteristic flow speed $\vphi$  to a global radial diffusion speed $v_{\rm diff,\infty}( \alpha,\varpi) = c/[3\tau_\infty(\alpha)]$, where $\tau_\infty( \alpha) = \int_{0}^\infty \kappa(\alpha,\varpi') \rho(\alpha,\varpi') d\varpi'$ is the optical depth to infinity at angle $\alpha$ from the breakout point. If $v>v_{\rm diff,\infty}$ along a streamline, we expect its radiation to be trapped, but if $v_{\rm diff,\infty} > v$ then it may escape to an external observer.   (Transfer across streamlines eases this criterion somewhat, by allowing photons from regions with $v_{\rm diff,\infty}<v$ to diffuse into streamlines on which $v_{\rm diff,\infty}>v$; however this is only likely where $\Diffusion <1$.)  The $\tau_\infty$ integral diverges logarithmically at large $\varpi'$ in two-dimensional constant-velocity flow,  and must be truncated at $\varpi\simeq R_*$ because three-dimensional effects set in on the scale of the stellar radius.  We conduct the radial integration outward from the breakout point using our numerical results, and then extrapolate to $R_\star$ assuming $\rho\propto \varpi^{-1}$ in each direction.  If we assume $R_*=100 \Lphi$ the resulting $\vphi/v_{\rm diff}(\alpha)$ is virtually identical to $3\Diffusion(\alpha)$, so
\begin{equation} \label{tauInfty} 
{\vphi \over v_{\rm diff,\infty}(\alpha_f) }  \simeq 3 \Diffusion(\alpha_f) \ln\left(R_*\over 100\,\Lphi\right), 
\end{equation} 
and because $\Diffusion\propto \alpha_f^{-6}$ the angle at which $\Diffusion = 1$ is only 1.2 times smaller than the angle at which $v/v_{\rm diff,\infty}=1$ (provided there is such an angle).  Practically speaking, this means that the condition for photon diffusion to affect the flow  ($\Diffusion<1$) is only slightly less restrictive than the condition for photons to escape the system entirely ($\vphi < v_{\rm diff,\infty}$). 

To evaluate the importance of diffusion in real stars, let us start by assuming (as we have in the simulations) that $\Lphi\ll R_*$, so that the zone of oblique flow is in a thin outer layer where $\rho_0 = \rho_h [-y/R_*]^n$. The shock velocity in this zone, neglecting non-radial motions, is $\hat v_s = C_1 v_* [M_{\rm ej}/(R_*^3 \rho)]^\beta$, if $C_1= 0.794$ and $\beta= 0.18575$ are parameters in the theory developed by \citet{1999ApJ...510..379M}, which can both be refined to account for details of the stellar structure \citep[][see also eqs.~(\ref{eq:beta-from-n}) and (\ref{eq:vfvs-from-n})]{2001ApJ...551..946T,2013ApJ...773...79R}.  The characteristic shock velocity $v_* = (E_{\rm in}/M_{\rm ej})^{1/2}$ is independent of location in a spherical explosion and nearly constant in the cases we consider here.  Identifying the depth $|y|=\Lphi$ where $\hat v_s = \vphi$, 
\[ \rho_\varphi = \left(\ C_1 v_*\over \vphi \right)^{1/\beta} {M_{\rm ej}\over R_*^{3}}\] 
and
\[ \Lphi = R_* \left(\rho_\varphi\over \rho_h\right)^{1/n};  \] 
therefore, with equation (\ref{eq:Diffusion}), 
\begin{equation}\label{eq:DiffEvaluated} 
\Diffusion \simeq {D_1\,\tau_h\over \alpha_f^{\delta_{\Diffusion}} } \left(M_{\rm ej}\over R_*^3 \rho_h\right)^{\gamma_p} 
\left( C_1 v_*\over \vphi \right)^{\gamma_p/\beta -1}
 {C_1 v_*\over c}
\end{equation} 
where $\tau_h = \kappa \rho_h R_*$ is a characteristic optical depth of the envelope, and $\gamma_p=1+1/n$ is the polytropic index.  

Our simulations only provide information about the case $\gamma_p=4/3$, but we can extrapolate to other polytropic indices by appealing to the fact that our numerical fit to $\Diffusion(\alpha_f)$ is quite close to what we would have obtained by extrapolating from the deep, planar flow.   In this deep-flow limit ($\alpha_f\rightarrow 0$),  we find in the Appendix that 
\begin{equation}\label{eq:DiffFromDeepFlow}
D_1 \rightarrow {9\over 4+n(3-2\beta)} C_2^d; ~~~{\delta_{\Diffusion}} \rightarrow {\gamma_p\over \beta} 
\end{equation} 
and $n$, $C_2$, and $\beta$ are all functions of $\gamma_p$. 
It is therefore reasonable to scale the numerical fits for $\Diffusion$ and ${\delta_{\Diffusion}} $ by the factors $\Diffusion(\gamma_p)/\Diffusion(4/3)$ and ${\delta_{\Diffusion}} (\gamma_p)/{\delta_{\Diffusion}} (4/3)$, respectively.  Although we do not yet know the accuracy of this extrapolation for significantly different $\gamma_p$, we adopt this approach in Table \ref{Table:Examples}.

In Table \ref{Table:Examples} we use the properties of the model core-collapse explosions introduced in Paper 1 to identify $\alpha_\Diffusion$,  the critical angle at which $\Diffusion=1$, as well as the depth ratio $R/\Lphi$ and the relativity factor $c/(2\vphi)$, in terms ${\cal V}_\varphi = \vphi t_{\rm se}/R_*$. We use this particular combination because, in the toy model of an aspherical explosion introduced by Paper 1,  ${\cal V}_\varphi$ is a known function of latitude $\theta$ if the explosion is elongated by the factor $\varepsilon$ at breakout: ${\cal V}_\varphi = 1/[2 \varepsilon \sin(2\theta)]$.  

The last three columns of Table 1 allow us to estimate the range of ${\cal V}_\varphi$ over which the assumptions of our numerical models hold: initial plane symmetry ($\Lphi\ll R$), non-relativistic flow $c \gg 2 \vphi$, and adiabatic flow ($\alpha_d \gg \pi/2$).  In all cases the lower limit, ${\cal V}_\varphi\gtrsim 0.6,$ comes the requirement of plane symmetry.  In extended stars, diffusion sets the upper limit (${\cal V}_\varphi \lesssim 2.2$ and ${\cal V}_\varphi \lesssim 6.4$ in red and blue supergiants, respectively), but in the compact Ic progenitor the upper limit is set by relativity (${\cal V}_\varphi \lesssim 21$ and ${\cal V}_\varphi \lesssim 3.8$ for $E_{51}=1$ and $E_{51}=30$, respectively), because shock propagation continues to relativistic speeds in such stars.  While our results are barely applicable to explosions in diffuse stars like red supergiants (for which diffusion sets in at a significant depth), they are relevant for more compact progenitors so long as the explosion is not too relativistic. 

Finally, we note that the maximum post-shock pressure is reached at the location of the kink, where the shock velocity is $v_s=\vphi$ and therefore $P_2  =(6/7) \rho_0 \vphi^2 \simeq \rho_\varphi\vphi^2$. This corresponds to a maximum blackbody  temperature in local thermodynamic equilibrium:  $T_\varphi = (3 \rho_\phi \vphi^2/a)^{1/4} \simeq  \{61{\cal V}_\varphi^{-0.80}\,\rm{eV}, 270 {\cal V}_\varphi^{-0.86}\,{\rm eV}, 18.2{\cal V}_\varphi^{-0.87}\,\rm{keV}\}\times E_{51}^{1/4}$~K for our model \{RSG, BSG, Ic\} progenitors.    Observed photons may be significantly more energetic if their population falls short of LTE \citep{2010ApJ...716..781K}, or they may be nearly an order of magnitude lower if they are degraded adiabatically (to the pressures found along the stellar surface) before escaping.  We leave this and other questions, such as the possibility of bulk Comptonization in the zone of ejecta-star shear flow, for later investigation. 

\section{Energy in non-radial motions}\label{S:TransverseEnergy} 

Some of the energy diverted by the oblique shock into non-radial flow will be available to power transients from collisions outside the stellar surface.  To calculate this, we note that the component of velocity parallel to the stellar surface in the star's frame ($v_{fx}'$, in our terminology) equals $-[1-\cos(\alpha_f)]$ at large distances.  The non-radial component of the energy flux is therefore $\rho \vphi^3 [1-\cos(\alpha_f)]^2$.  Computing the non-radial energy per unit stellar surface area for all streamlines diverted more than $\alpha_f$ in the shock's frame (or $\alpha_f/2$ in the star's frame), we find 
\begin{equation} 
{dE_x (>\alpha_f) \over dA} = {\rho_\varphi \vphi^2 \Lphi \over 2} \int_{\alpha_f}^\pi {\rho(\alpha_f') \varpi\over \rho_\varphi \Lphi} \left[1-\cos(\alpha_f')\right]^2 d\alpha_f'. 
\end{equation} 
This integral diverges in the limit $\alpha_f\rightarrow 0$, because $\rho \varpi \propto \alpha_f'^{-(1+\gamma_p/\beta)}\sim \alpha_f'^{-8.2}$ whereas $[1-\cos(\alpha_f')]^2 \propto \alpha_f'^4$.   However, the energy in {\em significantly} deflected motions -- deflected by at least $45^\circ$ in the star's frame  ($\delta \theta_f > \pi/4$) -- is about $0.51 \rho_\varphi \vphi^2 \Lphi$ per unit area. 
This tends to strongly weight those regions of the stellar surface where the pattern speed is lowest, because $\rho_\varphi \vphi^2 \Lphi\propto \vphi^{-(\gamma_p/\beta -2)} v_*^2 \sim  \vphi^{-5.2}v_*^2$.  
Integrating over the surface of the bipolar explosion model introduced in Paper 1, we find that a fraction $\left\{0.25\varepsilon^{4.5}, 0.33\varepsilon^{4.8}\right\}$ of the explosion energy is channeled into such motions in our \{BSG, Ic\} model progenitors, so long as the conditions for planar, non-relativistic flow are all met. 

\section{The effect of gravity} \label{S:Gravity}

In many of the astrophysical circumstances where oblique shock breakouts should occur, gravity is not guaranteed to be a negligible perturbation to the dynamics.  Examples include the ejection of planetary atmospheres during planetary collisions and giant impacts \citep{2003Icar..164..149G}; stellar collisions \citep{2012ApJ...759...39H};  eruptions of luminous blue variable stars, such as $\eta$ Carinae \citep{2013MNRAS.429.2366S}, accretion-induced collapses of white dwarfs \citep{1999ApJ...516..892F,2001ApJ...551..946T}, detonations on neutron stars \citep{2001ApJS..133..195Z,2012ApJ...755....4T}, and compression shocks in stellar tidal disruption events \citep{2009ApJ...705..844G}. 

To be specific, let us consider shock ejection from a spherical star with radius $R_*$, escape speed $\vesc$, and surface gravity $g=\vesc^2/(2R_*)$.   The ratio between gravity and the characteristic acceleration of an oblique shock breakout is $g\Lphi/\vphi^2 = 2(R_*/\Lphi)(\vphi/\vesc)^2$.   Therefore, so long as the explosion is nearly spherical ($R_*\gg \Lphi$), there exists a broad range of $\vphi$ for which gravity is a small perturbation on the scales of our simulation ($g\Lphi/\vphi^2\ll 1$), but gravity is nevertheless important on scales of order $R_*$ because $\vphi\lesssim \vesc$.   Under these conditions we can extrapolate the influence of gravity from our current results, much as we did for radiation diffusion in \S~\ref{S:OpticalDepth}.  

First, what is the condition for matter to be ejected?   Matter be accelerated to $\vesc$ in the rest frame of the star; since the maximum speed in that frame is $2\vphi$, a necessary condition for escape is $\vphi > \vesc/2$.  However, while this condition guarantees that matter will escape for a brief period,  it is probably not, in fact, sufficient for the ejection of matter in steady state.  The reason is that any streamline along which matter is bound to the star must fall back, and its trajectory can intersect that of matter ejected at $\vesc$ in the star's frame.   In particular, matter cast forward of the breakout region in the shock's frame ($\alpha_f>\pi/2$) may rain back on that region.  Furthermore, since the mass flow per unit angle decreases rapidly with $\alpha_f$, the ram pressure of this returning matter may overwhelm that of the otherwise escaping streamlines.  Matter with $\alpha_f=\pi/2$ achieves a speed $\sqrt{2} \vphi$ in the star's frame, so we estimate the criterion for mass ejection to be 
\begin{equation} \label{eq:escape} 
\vphi \gtrsim \frac{\vesc}{\sqrt{2}};
\end{equation}
i.e., the patten speed must exceed the surface Kepler velocity $(gR_*)^{1/2}$.  

This planar result is little more than an educated guess, however, for several reasons.  The interaction between out-flowing and returning matter is guaranteed to be complicated.  Spherical and non-steady effects will always be important for matter ejected at speeds of order $\vesc$.  Moreover, $\vesc$ itself could change if the explosion lifts enough material to change the gravitational potential \citep[e.g.][]{2004MNRAS.354.1053W}.   

If condition (\ref{eq:escape}) is not met, matter cast vertically at speed $\vphi$ (in the shock's frame) reaches a height $\vphi^2/(2g)$ before raining back.  This and the nearby streamlines will return to collide with the matter emerging from the breakout region. 

Second, how does gravity  affect the deep flow when $\vphi> \vesc$? Equation (\ref{eq:v_star_frame}) indicates that matter deflected by a sufficiently small amount, $\alpha_f < 2 \sin^{-1} [\vesc/(2 \vphi)]$, moves below the escape velocity in the star's frame.  However, as  spherical effects may be important at this depth, and as this matter is not in a region of strongly non-radial flow, the division between capture and escape is better described within a spherical theory (e.g., \S 2.6 of \citealt{2001ApJ...551..946T}). 

\section{Conclusions} \label{S:conclusions}

We have presented high resolution, two-dimensional simulations of oblique supernova shock breakout, focusing on the limit of adiabatic, non-relativistic flow in a thin layer near the stellar surface in order to compare our results against the analytical predictions and suggestions of Paper 1.   As expected, we find that the primary shock curves toward the stellar surface, experiences a kink, and reaches the surface almost vertically. The arrival of the shock at the surface is accompanied by a spray of matter ejected at a range of angles, including along the stellar surface.  The stellar surface is compressed by ejecta pressure, so that the breakout point is inside the original stellar radius.  The post-shock flow around the breakout point is neither steady nor self-similar: it exhibits acoustic oscillations, entropy fluctuations, and vortices. We study the effects of grid resolution and box size on the ejecta behavior, and show that the waves and vortices around the breakout point are independent of box size and only weakly dependent on resolution.   

Our simulations assume radiation diffusion, relativistic effects, gravity, and stellar curvature are all negligible, yet provide some guidance for scenarios in which they are not.  Diffusion, in particular, is more important for the shallowest (most strongly-deflected) material than for the deepest matter; we are therefore able to define a deflection angle $\alpha_{\cal D}$ above which diffusion is important and radiation may escape to be observed.  Radiation often is completely trapped ($\alpha_{\cal D} > \pi$) in compact stars, as predicted by Paper 1.   Gravitational effects are negligible for the strongly non-radial portions of the flow when the lateral pattern speed is well above the stellar escape velocity, but outflow is quenched by returning matter when $\vphi$ is approximately the stellar Kepler velocity. 

The vortices we observe in our strictly two-dimensional simulations are undoubtedly more pronounced than they would be three dimensions, thanks to the well-known inverse nature of 2D turbulence cascades \citep{1967PhFl...10.1417K}.  As these vortices are superimposed on a rapidly expanding flow, we see little to suggest that three-dimensional results would be qualitatively different; however this must be checked with future simulations. 

While the simulations presented here exploit the local, essentially two-dimensional nature of the problem to obtain very high effective resolution (up to 16,384$^2$), this is usually not possible in global simulations.  In such simulations, the effect of limited resolution on the structure of non-radial flows will be critically dependent on our resolution parameter $\Resolution$, i.e., the comparison between the local obliquity scale $\Lphi$ and the minimum grid spacing $\Delta x_{\rm min}$.  If this ratio is of order unity or less, we anticipate that the maximum possible deflection is limited to the small value appropriate for matter at the minimum resolvable depth: in the star's frame, this is an angle of approximately $\Resolution^{n/5}$ if the polytropic index is $n$ (equations [\ref{eq:alphaf_thetaf_relation}] and [\ref{eq:deep_limit_y0}], taking $\beta\simeq 1/5$ and $C_2\simeq 2$).   In other words, insufficient resolution should suppress the development of non-radial flows much as radiation diffusion does within extended stellar progenitors. 

One might question whether simulations which assume perfectly trapped radiation can be relevant to observations, so we end by reviewing why they are.  When radiation is truly trapped, as in aspherical type I supernovae, non-radial flows eliminate the shock breakout emission, which was expected in spherical theory, over much of the stellar surface.  By deflecting ejecta, non-radial motions allow collisions outside the star which may give rise to a novel form of transient.  For these reasons, and because the ejecta speeds are limited, circumstellar interactions and the early supernova light curve are altered.   Our local simulations provide quantitative estimates for the output of each patch of the stellar surface, which can be integrated to give global estimates  as we have done in \S \ref{S:TransverseEnergy}.  

When radiation is only partially trapped on the scales of interest, as in the explosions of blue supergiants, our results provide a way to estimate the critical angle above which radiation will escape ($\alpha_{\cal D}$ in \S~\ref{S:OpticalDepth}), which we also expect to be the limiting deflection angle.  We anticipate that the emerging luminosity will match the kinetic luminosity above $\alpha_{\cal D}$   in our adiabatic simulations, but radiation hydrodynamic simulations will be required to test this. 

Finally, when radiation is poorly trapped ($\alpha_{\cal D} \ll 1$), as in the explosions of extended red supergiant stars, our simulations are not relevant: the dynamics of shock breakout are described, at least locally, by the theory for spherical explosions. 

\acknowledgements
PS, CDM, and SR are supported by an NSERC Discovery Grant; YL's research is supported by an ARC Future Fellowship. We thank Ian Parrish and Shane Davis for scientific discussions and extremely useful help with Athena, Kristen Menou for bringing to our attention the relevance of this theory to planetary collisions, and Nathan Smith and John Bally for pointing out the possible relevance to $\eta$ Car.  We are also very grateful to James Guillochon for insightful comments and to Chris McKee and the referee for stimulating questions. 
Most of the computations were performed on the GPC supercomputer at the SciNet HPC Consortium. SciNet is funded by the Canada Foundation for Innovation under the auspices of Compute Canada; the Government of Ontario; Ontario Research Fund - Research Excellence; and the University of Toronto.

\appendix

\section{Deep flow,  small-deflection limit of ejecta distribution} \label{Appendix:DeepFlowLimit}

Within our idealization that $\Lphi$ is very much less than the stellar radius (and any other scale over which shock properties vary), the very deep flow ($|y_0|\gg \Lphi$) is governed by the planar, self-similar solution to shock acceleration and post-shock flow.   As a result, we can use the solutions described by \citet{sakurai60} and \citet{1999ApJ...510..379M} to determine the angular dependence of all the fluid variables which emerge from this deep region, i.e., the shallow-angle flow ($\alpha_f \ll1$).  

In the deep flow where corrections due to non-radial motions are negligible, our definitions and the shock acceleration law imply $v_s/\vphi = [\rho_\varphi/\rho_0(y_0)]^\beta = (\Lphi/|y_0|)^{n\beta}$.  In planar flow each fluid element reaches a terminal vertical (radial) velocity $v_f(y_0) = C_2 v_s(y_0)$; given a total speed $\vphi$ in the shock frame (a consequence of energy conservation), this implies a final angle $\alpha_f = C_2 v_s/\vphi$ in the small-angle limit.   

In this limit each flow quantity $F$ will tend far downstream ($\alpha_f \varpi_f \gg |y_0|$) toward a power-law form 
\[ F(\varpi,\alpha_f) \rightarrow {K_F F_\varphi \over \alpha_f^{\delta_F} } \left( \Lphi\over \varpi\right)^{k_F} \] 
where $F_\varphi$ is the combination of $\vphi$, $\rho_\varphi$, and $\Lphi$ of the same dimensions as $F$, and $K_F$, $\delta_F$, and $k_F$ are dimensionless constants.  So, for instance, the relation $y_0 = - \Lphi (C_2/\alpha_f)^{1/(n\beta)}$ can be expressed 
\begin{equation} \label{eq:deep_limit_y0} 
K_{y_0} = C_2^{\delta_{y_0}},  ~~~~~~  \delta_{y_0}={1\over n\beta},  ~~~~~~ k_{y_0}=0. 
\end{equation}
Mass conservation requires $\rho(\alpha_f,\varpi) \varpi = \rho_0(y_0) |dy_0/d\alpha_f|$, so 
\begin{equation} \label{eq:deep_limit_rho}
K_\rho = { C_2^{\gamma_p/\beta } \over n\beta}, ~~~~~~ \delta_\rho ={\gamma_p\over \beta}+1,  ~~~~~~ k_\rho = 1
\end{equation} 
The post-shock entropy is $s(y_0) = P_2(y_0)/\rho_2(y_0)^\gamma$, where $P_2(y_0) = 2\rho_0(y_0) v_s(y_0)^2/(\gamma+1)$ and $\rho_2(y_0) = (\gamma+1)\rho_0(y_0)/(\gamma-1)$; therefore 
\begin{equation} \label{eq:deep_limit_s}
K_s  = {2\over \gamma+1}\left(\gamma-1\over\gamma+1\right)^\gamma C_2^{\delta_s},  ~~~~~~ \delta_s = -\left({\gamma-1\over \beta} + 2\right),  ~~~~~~ k_s = 0.  
\end{equation} 
The pressure is $P = s\rho^\gamma$ so
 \begin{equation}\label{eq:deep_limit_P}
 K_P = K_s K_\rho^\gamma, ~~~~~~ \delta_P = \delta_s + \gamma\delta_\rho = {\gamma/n+1\over\beta}+\gamma-2,  ~~~~~~ k_P = \gamma.
 \end{equation} 
 To construct the pressure scale length ${\cal L}_p = P/|\nabla P|$ consider that $\nabla P = (\partial P/\partial \alpha) \hat \alpha/\varpi + (\partial P/\partial \varpi) \hat \varpi \rightarrow - (P/\varpi) [(\delta_p/\alpha) \hat \alpha + \gamma \hat \varpi]$ (where $\hat \alpha, \hat \varpi$ are unit vectors).  The first term dominates for $\alpha\ll1$, so 
  \[ {\cal L}_p(\alpha_f, \varpi) \rightarrow {\alpha_f \varpi\over \delta_P}.\]   
  The local diffusion parameter $\Diffusion$ from \S\,\ref{S:OpticalDepth} is defined as $3 \kappa \rho {\cal L}_p v/c$.  Because $v\rightarrow \vphi$, 
 \begin{equation} \label{eq:deep_limit_Diffusion}
 K_{\Diffusion} = {3\kappa \rho_\varphi \vphi  \over \beta n  c \,\delta_P}C_2^{\delta_{\Diffusion}}, ~~~~~~ \delta_{\Diffusion} = {\gamma_p\over \beta}, ~~~~~~ k_{\Diffusion} = 0. 
 \end{equation}  
Written out with $\gamma=4/3$, this yields the expressions given in equation (\ref{eq:Diffusion}) with the constants in equation (\ref{eq:DiffFromDeepFlow}). 

Finally we wish to provide fits for $\beta(n)$ and $C_2(n)$ which, being specialized to the case $\gamma=4/3$, are simpler than, and comparably accurate to, those given by \citet{2013ApJ...773...79R}: these are
\begin{equation} \label{eq:beta-from-n}
\beta = \beta_0 + {\beta_\infty - \beta_0\over (1.33281/n)^{1.01595}+1 }
\end{equation} 
 where $\beta_0 = 1/(2+\sqrt{8})$ and $\beta_\infty = 0.17639782$; and 
 \begin{equation} \label{eq:vfvs-from-n}
C_2 = C_{2,0} + {C_{2,\infty}-C_{2,0} \over ( 0.311/n)^{0.822} + 1}
 \end{equation} 
 where $C_{2,0} = 6(1+\sqrt{8})/7$ and $C_{2,\infty} = 1.83941$.  Equation (\ref{eq:beta-from-n}) has an r.m.s. absolute error of $0.8\times 10^{-5}$, and equation (\ref{eq:vfvs-from-n}) has an r.m.s. relative error of 0.35\%, when compared against the numerical solutions obtained by \citeauthor{2013ApJ...773...79R}.
 
\bibliographystyle{apj}
\bibliography{SNGRB}
 
\end{document}